\documentclass[aps,prb,superscriptaddress,groupedaddress,floatfix,twocolumn,showpacs]{revtex4}
\usepackage{graphicx}
\usepackage{dcolumn}
\usepackage{bm}
\usepackage{bbold}
\usepackage{stmaryrd}
\usepackage{latexsym}
\usepackage{amssymb}
\usepackage{amsfonts}
\usepackage{amsmath}
\usepackage{mathtools}
\usepackage{fancybox}
\usepackage{color}
\usepackage[breaklinks=true,colorlinks,citecolor=blue,linkcolor=blue,urlcolor=blue]{hyperref}

\begin{document}
\title{Quasiparticle Mass Enhancement and Fermi Surface Shape Modification \\  
in Oxide Two-Dimensional Electron Gases}
\author{John R. Tolsma}
\email{tolsma@physics.utexas.edu}
\affiliation{Department of Physics, The University of Texas at Austin, Austin Texas 78712, USA}
\author{Alessandro Principi}
\affiliation{Radboud University of Nijmegen, Institute for Molecules and Materials, Heijendaalseweg 135, 6525 AJ Nijmegen, The Netherlands}
\author{Reza Asgari}
\affiliation{School of Physics, Institute for Research in Fundamental Sciences (IPM), Tehran 19395-5531, Iran}
\author{Marco Polini}
\affiliation{Istituto Italiano di Tecnologia, Graphene Labs, Via Morego 30, I-16163 Genova,~Italy}
\author{Allan H. MacDonald}
\affiliation{Department of Physics, The University of Texas at Austin, Austin Texas 78712, USA}
\date{\today}
\begin{abstract}
We propose a model intended to qualitatively capture the electron-electron 
interaction physics of two-dimensional electron gases formed near transition-metal oxide heterojunctions
containing $t_{2g}$ electrons with a density much smaller than one electron per metal atom. 
Two-dimensional electron systems of this type can be described perturbatively
using a $GW$ approximation which predicts that Coulomb interactions enhance quasiparticle effective masses
more strongly than in simple two-dimensional electron gases,
and that they reshape the Fermi surface, reducing its anisotropy.
\end{abstract}
\maketitle

\section{Introduction}
\label{sect:Intro}
Transition-metal oxides in three dimensions display an amazing variety of novel 
phenomena, from high-temperature superconductivity and colossal magnetoresistance to orbital 
ordering and metal-insulator phase transitions.  Because the metal $d$-bands present near their 
Fermi levels tend to be narrow and sensitive to oxygen coordination, both electron-electron and 
electron-lattice interactions are often strong.  
When the number of $d$-electrons per transition metal site is close to an integer, the most important 
electron-electron interactions occur on the atomic length scale and can be captured by 
Hubbard-type model interactions~\cite{Hubbard}. 
In $d$-band systems it is also often important  
to distinguish the manner in which $d$-orbitals form bonds with neighboring oxygen ions~\cite{Mattheiss}.
These two features provide a framework for analyzing many strongly interacting bulk
transition-metal oxide crystals. 

It has recently~\cite{stemmer_apl_2013,hwang_apl_2010,mannhart_science_1010,stemmer_armr_2014,levy_armr_2014} become possible to realize two-dimensional quantum wells based on 
heterojunctions between transition-metal oxides, and while the nature of d-electron bonding remains important, Hubbard-like correlations are often not.  To date the most common quantum 
well material is SrTiO$_3$ and the number of $d$-electrons per metal in the quantum wells 
is typically, although not always\cite{stemmer_prb_2012,balents_prb_2013}, much less than one.  
Oxide two-dimensional electron systems have application potential because,
as in the case of covalent semiconductors, 
large relative changes in the quantum well carrier density can be achieved by electrical means.  
The conduction bands of these systems are formed from $t_{2g}$ electrons that  
are weakly $\pi$-bonded to neighboring oxygens, and consequently form rather 
narrow and anisotropic bands.  When the electron density per metal atom is 
much smaller than one, the Fermi surface occupies a small fraction of the Brillouin zone and 
the probability of two electrons simultaneously occupying the same transition-metal site is small.
In this limit, including only the Hubbard part of the full electron-electron interaction misses
the most important Coulomb interactions. Because of its long range the typical Coulomb
interaction energy of an individual electron drops to zero only as two-dimensional density $n^{1/2}$, 
in contrast to the $\propto n$ behavior of the Hubbard model.
Indeed, the full long-range of the Coulomb potential must be recognized in any theory 
of electron-electron interaction effects in semimetal or doped semiconductor
small-Fermi-surface systems.   
The long-range Coulomb potential plays a critical role in the theory of plasmon 
oscillations~\cite{Bohm_and_Pines,jonson,stern_prl_1967,grecu}, quasiparticle effective mass~\cite{vinter_prl_1975,santoro_prb_1989,Asgari_PRB_2005}, angle-resolved photoemission spectra~\cite{bostwick_science_2010,polini_prb_2008}, and many other observables~\cite{Pines_and_Nozieres,Giuliani_and_Vignale}. 

In this Article we introduce a generic model for two-dimensional 
$t_{2g}$ electron gases which captures both the anisotropic character of the d-orbitals forming the low-energy
 conduction bands, and the importance of long-range Coulomb interactions when the number of conduction electrons per
 transition-metal site is much less than one. The $t_{2g}$ two-dimensional electron gas model we introduce
 is informed by recent self-consistent Hartree/tight-binding~\cite{guru_prb_2012,millis_prb_2013} 
and \emph{ab initio} calculations~\cite{ghosez_prl_2011,guru_prb_2013,Popovic_prl_2008} for SrTiO$_3$ quantum wells.  As a first application of this model, we calculate some observable quasiparticle properties of electrons in the 
anisotropic bands. Our Article is organized as follows. 
In Section~\ref{Sect:Model} we describe the $t_{2g}$ two-dimensional electron gas model and discuss the limits of its validity as a model of SrTiO$_3$ quantum wells. 
In Section~\ref{Sect:Self_Energy} we describe the $G_0W$ approximation for the quasiparticle self-energy and present 
explicit expressions for its {\it line-residue} decomposition~\cite{Quinn_and_Ferrell}. 
We use these expressions to calculate the renormalized Fermi surface shape in Section~\ref{Sect:FSrenorm} and the quasiparticle mass enhancement in Section~\ref{Sect:Mass}. 
Finally, in Section~\ref{Sect:Summary} we present our conclusions. 
\section{The $t_{2g}$ Two-Dimensional Electron Gas Model}
\label{Sect:Model}
\begin{figure}[t]
\includegraphics[width=0.95\linewidth]{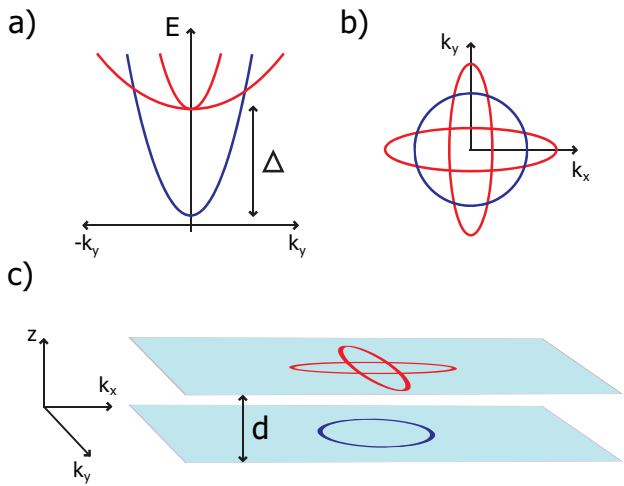}
\caption{(Color online) A schematic summary of the $t_{2g}$ two-dimensional electron gas model. 
In a) the energy offset $\Delta$ between the anisotropic $xz$ and $yz$ (red) band edge and the isotropic $xy$ (blue) band edge is emphasized. In b) the anisotropy of the elliptical $xz$ and $yz$ Fermi surfaces (red) is contrasted with the $xy$ band's circular Fermi surface (blue). The lower panel highlights the difference in $\hat{\bm z}$-direction confinement between the $xz$ and $yz$ 
bands and the $xy$ band, which can be crudely characterized by a separation distance $d$. \label{fig:zero}}
\end{figure}
The many-body effects of long-range Coulomb interactions in 
covalent semiconductors are often studied using continuum electron gas 
models~\cite{Pines_and_Nozieres,Giuliani_and_Vignale}.  In order to apply a similar approach to  
the two-dimensional electron gas residing at heterojunctions between SrTiO$_3$ and a barrier material,
it is necessary to account for some key differences.  These are captured by the $t_{2g}$ two-dimensional electron 
gas ($t_{2g}$ 2DEG) model which we now detail. The three distinct characteristics 
of $t_{2g}$ 2DEGs are band-mass anisotropy, a energy offset between the band edges of
subbands with different masses along the confinement direction, 
and a band-dependence in the distance between two-dimensional subband density maxima  
and the heterojunction or surface. Figure~\ref{fig:zero} illustrates these three features.
The model parameters we choose for the illustrative calculations we describe later in this Article 
are informed by recent tight-binding~\cite{guru_prb_2012,millis_prb_2013} and \emph{ab initio} 
calculations~\cite{ghosez_prl_2011,guru_prb_2013,Popovic_prl_2008} for SrTiO$_3$ 2DEGs. 
The $t_{2g}$ 2DEG model can be adapted to other materials by adjusting the model parameter choices.

The anisotropic nearest-neighbor effective metal-to-metal hopping amplitudes of $t_{2g}$ orbitals 
are ultimately responsible for many of the distinct characteristics of the $t_{2g}$ 2DEG. 
In three-dimensional SrTiO$_3$, crystal-field splitting breaks the 5-fold degeneracy amongst the Ti d-orbitals. 
The $xy$, $yz$, and $xz$ orbitals ({\it i.e.} the $t_{2g}$ orbitals) are lower in energy because they bond 
less strongly with neighboring oxygens and form the bulk conduction bands in n-type semiconducting SrTiO$_3$.
Electronic hopping between $t_{2g}$ d-orbitals on different Ti sites proceeds in a two-step process via the 
octahedrally coordinated oxygen p-orbitals surrounding each Ti atom~\cite{guru_prb_2012,guru_prb_2013,Mattheiss,Goodenough}. Orbital symmetry dictates that each $t_{2g}$ electron hops mainly between
states with the same orbital character, with a large hopping amplitude $t$ in two directions, and a 
smaller hopping amplitude $t^{\prime} $ in the third. 
This leads to separate $t_{2g}$ bands with $xy $, $yz $, and $xz $ d-orbital character that have 
a light mass $m_{\rm L}$ in two directions, and a heavy mass $m_{\rm H}$ in the third direction.  
The weak hopping directions for $xy$, $xz$, and $yz$ are  
$\hat{\bm z}$, $\hat{\bm y}$, and $\hat{\bm x}$, respectively.
The heavy masses are therefore in perpendicular directions for the three orbital characters.
In the presence of a $\hat{\bm z}$-direction confining potential strong enough to produce an effectively two-dimensional 
system, the $xy$ band has two light masses in-plane, while the $xz$ and $yz$ bands have one heavy
and one light mass in-plane (see Fig.~\ref{fig:zero}). According to recent tight-binding fits~\cite{guru_prb_2012} to Shubnikov-de Haas measurements~\cite{Allen_prb_2013} of bulk n-type SrTiO$_3$, $m_{\rm H} = \hbar^2/(2 t' a^2)$ 
and $m_{\rm L} = \hbar^2/(2 t a^2)$ where $t = 236~{\rm meV}$, 
$t' = 35~{\rm meV}$, and the lattice constant $a = 3.9$~\AA. 
This implies that  
\begin{equation}\label{eq:heavymass}
\frac{m_{\rm H}}{m} = \frac{{\rm Ry}}{t'}\left(\frac{a_{\rm B}}{a}\right)^2 \sim 7
\end{equation}
and
\begin{equation}\label{eq:lightmass}
\frac{m_{\rm L}}{m} = \frac{{\rm Ry}}{t}\left(\frac{a_{\rm B}}{a}\right)^2 \sim 1~,
\end{equation}
where $m$ is the bare electron mass in vacuum, 
$a_{\rm B} = \hbar^2/(m e^2) =0.529$~\AA~is the Bohr radius, and 
${\rm Ry} = \hbar^2/(2 m a^2_{\rm B}) = 13.6~{\rm eV}$ is the Rydberg energy.
These values for the heavy and light mass are in good agreement with angle-resolve photoemission spectrum measurements on bulk SrTiO$_3$~\cite{Rotenberg_prb_2010}.

As a result of these relatively large effective band-mass values and the non-linear and non-local 
dielectric screening properties of bulk SrTiO$_3$, subband splitting is relatively small and
several subbands of $xy$, $yz$, and $xz$ type are expected to be 
occupied even at moderate electron densities~\cite{guru_prb_2012,millis_prb_2013,Popovic_prl_2008}.
However, since even in this case $\gtrsim 75\%$ of the electron density is contained in
the lowest $xy$, $yz$ and $xz$ subband~\cite{guru_prb_2012}, in the $t_{2g}$ 2DEG model we 
address the case in which only one subband 
of each orbital type is occupied. This model is sufficiently realistic to account for the most interesting peculiarities 
of this type of 2DEG and can be generalized if there is interest in describing the properties of particular 
2DEG systems which have more occupied subbands.  
Figure~\ref{fig:zero} illustrates the anisotropy of the $xz$ and $yz$ bands in the three-band $t_{2g}$ 2DEG
model.

In addition to the band-mass anisotropy, two other
important characteristics of the $t_{2g}$ electron gas model follow from the anisotropic hopping amplitudes of 
the $t_{2g}$ d-orbitals. First, because $xz$ and $yz$ electrons have a much larger hopping amplitude in the $\hat{\bm z}$-direction (which we take to be the confinement direction) than 
the $xy$ electrons whose heavy mass is in the $\hat{\bm z}$-direction, the former are less easily confined to the
same surface or to an interface of a heterojunction system~\cite{guru_prb_2012,ghosez_prl_2011}.
This separation introduces an orbital dependence to the electron-electron interactions which we capture 
by introducing an effective distance $d$ between the $xy$ and the $xz$ and $yz$ bands.
Realistic values of $d$ in SrTiO$_3$ can be estimated
from previously published studies of the layer dependent $t_{2g}$ density distribution
as a function of confinement field~\cite{guru_prb_2012,millis_prb_2013}.
The effective separation decreases for increasing interfacial confinement field and total $t_{2g}$ density and lies in the range $d=2a$ - $10a$ for $t_{2g}$ electron densities between  
$2 \times 10^{13}~{\rm cm}^{-2}$ - $3 \times 10^{14}~{\rm cm}^{-2}$. 
Second, mass differences in the confinement direction leads to a finite energy offset $\Delta$ between 
the conduction band edges of the $xy$ and the $xz$-$yz$ bands, see Fig.~\ref{fig:zero}a). 
The band offset increases for larger confinement field (total $t_{2g}$ density) and in SrTiO$_3$ theory has proposed that  $\Delta = 10$ - $200~{\rm meV}$~\cite{guru_prb_2012}, in reasonable agreement with the range of values found in recent experiments~\cite{king_natcomm_2014,Cancellieri_prb_2014,ilani_natcomm_2014}.

Motivated by these three distinct characteristics of $t_{2g}$ 2DEGs and recognizing that it is 
necessary to account for the long-range Coulomb interaction, 
we propose the following Hamiltonian for the $t_{2g}$ 2DEG: 
\begin{equation}\label{eq:2DEGHamiltonian}
\begin{array}{l}
{\displaystyle {\cal H}_{t_{2g}}= \sum_{{\bm k} \alpha} \varepsilon_{\alpha}({\bm k}) \; \hat{c}^{\dagger}_{{\bm k} \alpha}\hat{c}_{{\bm k}\alpha}}\vspace{0.1 cm}\\
{\displaystyle + \frac{1}{2 A} \; \sum_{{\bm q} \neq 0} \sum_{\mathclap{\substack{ {\bm k} {\bm k}^{\prime} \\ \alpha \alpha^{\prime}}}} V_{\alpha \alpha^{\prime}}(q) \; c^{\dagger}_{{\bm k}+{\bm q} \alpha}  c^{\dagger}_{{\bm k}^{\prime}-{\bm q} \alpha^{\prime}} c_{{\bm k}^{\prime} \alpha^{\prime}} c_{{\bm k} \alpha} }~, 
\end{array}
\end{equation}
where $\alpha$ represents both spin and band-orbital quantum numbers, 
$A$ is the 2D sample area, and the Fourier transform of the 2D Coulomb interaction is 
\begin{equation}\label{eq:interactions}
V_{\alpha \alpha^{\prime}} \left( q \right)=\frac{2 \pi e^2}{\kappa q}e^{-q d_{\alpha \alpha^{\prime}}} \equiv v_qe^{-q d_{\alpha \alpha^{\prime}}}~,
\end{equation}
with $q = |\bm{q}|$. Here, $\kappa$ is an effective dielectric constant and 
$d_{\alpha \alpha^{\prime}}$ gives the effective confinement-direction separation distance between an electron with band index $\alpha$ and an electron with band index $\alpha^{\prime}$. For the typical electron-electron interaction transition energies in $t_{2g}$ electron gases the relevant dielectric constant does not include soft-phonon contributions~\cite{Comment}, but depends on the dielectric environment on both sides of the 
relevant heterojunction or surface. The $t_{2g}$ band energies near the band minimum are
\begin{equation}\label{eq:ellipticalbands}
\varepsilon_{\alpha}({\bm k}) = \left\{
\begin{array}{l}
{\displaystyle   \frac{\hbar^2 {\bm k}^2}{2 m_{\rm L}}\qquad \qquad  \qquad \quad \; \; {\rm for} \qquad \alpha=xy,\sigma}\vspace{0.1 cm}\\
{\displaystyle   \frac{\hbar^2 k^2_x}{2 m_{\rm L}} +\frac{\hbar^2 k^2_y}{2 m_{\rm H}} + \Delta \qquad  {\rm for} \qquad \alpha=xz,\sigma}\vspace{0.1 cm}\\
{\displaystyle \frac{\hbar^2 k^2_x}{2 m_{\rm H}} +\frac{\hbar^2 k^2_y}{2 m_{\rm L}} + \Delta \qquad {\rm for} \qquad \alpha=yz,\sigma}
\end{array}
\right.~,
\end{equation}
where $m_{\rm H}$ and $m_{\rm L}$ have been introduced earlier in Eqs.~(\ref{eq:heavymass}) and (\ref{eq:lightmass}), respectively.

The applicability of the proposed $t_{2g}$ 2DEG model (\ref{eq:2DEGHamiltonian}) to describe SrTiO$_3$-based 2DEGs depends on the total electron density.  The continuum model for the band structure is 
valid only if the number of conduction band electrons per Ti site is much smaller than one. 
Its applicability therefore depends not-only on the carrier density per cross-sectional area, but 
also on quantum well thickness.  
On the other hand, we have neglected spin-orbit coupling terms in the band Hamiltonian, which play an 
essential role at small carrier densities.  The $t_{2g}$ 2DEG model is applicable 
when the density is large enough that spin-orbit coupling, which acts to mix the orbital character of the conduction bands~\cite{guru_prb_2012}, can be neglected, {\it i.e.} when the strength of spin-orbit coupling 
({\it e.g.}~$\sim 17~{\rm meV}$ in Ref.~\onlinecite{Allen_prb_2013}) is small compared to the Fermi energy.  

\section{The quasiparticle self-energy in $G_0W$-RPA}
\label{Sect:Self_Energy}
Coulomb interactions in Fermi liquids, whether doped semiconductors or weakly correlated metals,
give rise to two types of elementary excitations~\cite{Pines_and_Nozieres,Giuliani_and_Vignale}: neutral collective excitations and charged quasiparticles. 
The latter, with which we are concerned in this Article, are excitations with the same quantum numbers as 
non-interacting independent-particle electronic states. Their energies are shifted from the non-interacting values
and their lifetimes are finite, in both cases because of electron-electron interactions. 

A self-consistent equation for the quasiparticle excitation energy,
{\it i.e.} the quasiparticle energy measured from the chemical potential, 
is obtained from the Dyson equation by locating the energies at which the 
spectral weight of the one-particle retarded Green's 
function~\cite{Giuliani_and_Vignale} is peaked:  
\begin{equation}\label{eq:QP_EE}
\mathcal{E}_{\alpha}({\bm k}) = \xi_{\alpha}({\bm k}) + {\rm Re} \bar{\Sigma}_{\alpha}({\bm k},\omega)\vert_{\omega = \mathcal{E}_{\alpha}({\bm k})/\hbar}~,
\end{equation}
where $ \xi_{\alpha}\left({\bm k}\right) = \varepsilon_{\alpha}({\bm k}) - \varepsilon_{\alpha}({\bm k}_{{\rm F} \alpha}) $ 
is the band energy measured relative to the Fermi energy,
and in the self-energy $\bar{\Sigma}_{\alpha}\left({\bm k},\omega \right) =  \Sigma_{\alpha}\left({\bm k},\omega \right) - \Sigma_{\alpha}\left({\bm k}_{{\rm F}\alpha},0 \right)$ we 
subtract the term  $\Sigma_{\alpha}\left({\bm k}_{{\rm F}\alpha},0 \right) $ to account for the interaction correction to the chemical potential $\mu$ given by
\begin{equation}\label{eq:chem_pot}
\mu = \varepsilon_{\alpha}({\bm k}_{{\rm F}\alpha}) + \Sigma_{\alpha}\left({\bm k}_{{\rm F}\alpha},0 \right)~.
\end{equation}
The real part of the self-energy in Eq.~(\ref{eq:QP_EE}) yields the many-body contribution to the energy of the quasiparticle state. 
The quasiparticle energy can be measured by taking angle-resolved photoemission spectra~\cite{bostwick_science_2010,polini_prb_2008}, and more indirectly by performing magneto-transport measurements~\cite{Fang_and_Stiles,pudalov_prl_2002,Tan_prl_2005}.
\begin{figure}[t]
\includegraphics[width=1.0\linewidth]{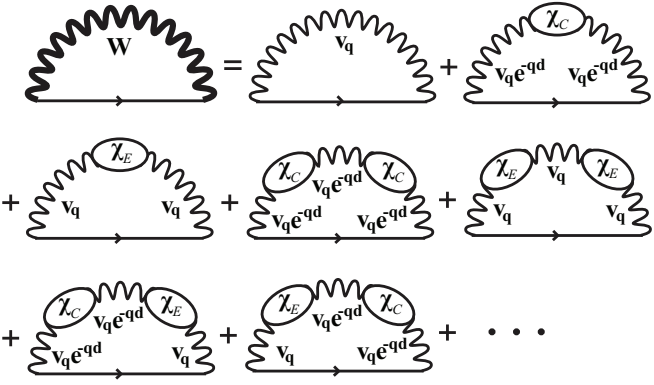}
\caption{Feynman diagrams representing the $G_0W$-RPA approximation for the self-energy of a quasiparticle in one of the elliptical bands of the $t_{2g}$ 2DEG. In RPA the bare ({\it i.e.} unscreened) Coulomb interaction is dressed ({\it i.e.} screened) by a infinite series of bubble diagrams representing the density fluctuations of each component ({\it e.g.} each band) of the system. When applied to the three-band $t_{2g}$ 2DEG, each bubble represents a density fluctuation in either the circular $xy$ band, $\chi_{\rm C} \equiv \chi^{(0)}_{xy}$, or the elliptical $yz$ and $xz$ bands, $\chi_{\rm E} \equiv \chi^{(0)}_{yz} + \chi^{(0)}_{xz} $. The circular $xy$ band and the elliptical $xz$ and $yz$ bands have different $\hat{\bm z}$-direction confinement, which we have crudely accounted for with the separation distance $d$. As a result, thin wavy lines representing the bare Coulomb interaction are equal to either $v_q$ or $v_qe^{-qd}$ depending on whether the density fluctuations attached to each thin wavy line's endpoints are in the same layer or different layers, respectively. Directed lines represent the non-interaction Green's function of the elliptical-band quasiparticle in question, as in Eq.~(\ref{eq:Greensfunction}) with $\alpha = yz~{\rm or}~xz$. The thick wavy line represents the fully screened RPA interaction given by Eq.~(\ref{eq:RPAinteraction}). \label{fig:Feynman}}
\end{figure}

The $G_0W$ approach, which we apply below, provides a 
successful\cite{Quinn_and_Ferrell,santoro_prb_1989,mahan_prl_1989,macdonald_prb_1994,dassarma_prb_2004,polini_prb_2008} 
approximation for the quasiparticle self-energy in electronic systems in which 
long-range Coulomb interactions play an essential role.  In the $G_0W$ approximation, we employ the random-phase approximation (RPA) for the screened electron-electron interaction. The screened interaction ${\bm W}$ (which is a $3\times 3$ matrix in the band indices $\alpha, \alpha^\prime$) is most simply derived by the following algebraic approach~\cite{Giuliani_and_Vignale}. A generalized Dyson equation relates ${\bm W}$ to the density-response matrix ${\bm \chi}$,
\begin{equation}\label{eq:screenedpotential}
{\bm W} = {\bm V} + {\bm V} \cdot {\bm \chi} \cdot {\bm V}~,
\end{equation}
where the matrix of unscreened Coulomb interactions is given by
\begin{equation}
{\bm V}(q) \equiv \left(
\begin{array}{ccc}
v_q & v_q e^{-qd} & v_q e^{-qd}\\
v_q e^{-qd} & v_q & v_q \\
v_q e^{-qd} & v_q & v_q
\end{array}
\right)~.
\end{equation}
The first, second, and third rows (columns) of each matrix in Eq.~(\ref{eq:screenedpotential}) correspond to the $xy$, $yz$, and $xz$ bands, respectively. In RPA the density-response matrix is related to the diagonal matrix of non-interacting density-response functions ${\bm \chi}^{(0)}({\bm q},\omega) \equiv {\rm diag}(\chi^{(0)}_{xy}({\bm q},\omega),\chi^{(0)}_{yz}({\bm q},\omega),\chi^{(0)}_{xz}({\bm q},\omega))$ via
\begin{equation}\label{eq:matrixform}
\big[ {\bm \chi}({\bm q},\omega)\big]^{-1} = \big[{\bm \chi}^{(0)}({\bm q},\omega)\big]^{-1} - {\bm V}({\bm q})~.
\end{equation}
Analytic expressions for ${\rm Re}\, \chi^{(0)}_{xy}({\bm q},\omega)$ and ${\rm Im}\, \chi^{(0)}_{xy}({\bm q},\omega)$ can be found, for example, in Ref.~\onlinecite{Giuliani_and_Vignale}. Expressions for the elliptical-band functions $\chi^{(0)}_{xz}({\bm q},\omega)$ [$\chi^{(0)}_{yz}({\bm q},\omega)$] can be easily found by applying the rescaling $k_x \to k_x \sqrt{m_{\rm L} /m_{\rm DOS}}$ and $k_y \to k_y \sqrt{m_{\rm H} / m_{\rm DOS}}$ [$k_x \to k_x \sqrt{m_{\rm H} / m_{\rm DOS}}$ and $k_y \to k_y\sqrt{m_{\rm L}/m_{\rm DOS}}$] in Eq.~(\ref{eq:ellipticalbands}), where $m_{\rm DOS} = \sqrt{m_{\rm L} m_{\rm H}}$ is the density-of-states mass. Since this rescaling maps the elliptical band onto a circular band, the elliptical band density-response functions $\chi^{(0)}_{xz}({\bm q},\omega)$ and $\chi^{(0)}_{yz}({\bm q},\omega)$ can be written in terms of $\chi^{(0)}_{\rm xy}({\bm q},\omega)$:
\begin{equation}\label{eq:polarizationxz}
\chi^{(0)}_{xz}({\bm q},\omega) =  \left. \chi^{(0)}_{\rm xy}(q',\omega; m_{\rm DOS}) \right|_{q' \to \sqrt{q^2_x \zeta^{1/2} + q^2_y \zeta^{-1/2}}}
\end{equation}
and
\begin{equation}\label{eq:polarizationyz}
\chi^{(0)}_{yz}({\bm q},\omega) =  \left. \chi^{(0)}_{\rm xy}(q',\omega; m_{\rm DOS}) \right|_{q' \to \sqrt{q^2_x \zeta^{-1/2} + q^2_y \zeta^{1/2}}}
\end{equation}
where we have defined $\zeta = m_{\rm H}/m_{\rm L}$. Eqs.~(\ref{eq:screenedpotential})-(\ref{eq:polarizationyz}) can be combined to yield analytic expressions for each element of ${\bm W}$. 

Later we will specifically be interested in the screened interaction between two electrons in the anisotropic $xz$ or $yz$ bands while in the presence of a circular $xy$-band Fermi sea. This particular interaction corresponds to the matrix element ${\bm W}_{xz,xz}$ (or equally ${\bm W}_{yz,yz}$) which we write as  
\begin{equation}\label{eq:RPAinteraction}
W_{xz}({\bm q},\omega) \equiv {\bm W}_{xz,xz}  = \frac{v_q + (e^{-2 q d}-1)v_q^2\, \chi^{(0)}_{xy}({\bm q},\omega)}{\varepsilon({\bm q},\omega)}~,
\end{equation}
where the RPA dielectric function is given by
\begin{eqnarray}\label{eq:RPAdielectric}
\varepsilon({\bm q},\omega) & = & [1-v_q\chi^{(0)}_{xy}][1-v_q(\chi^{(0)}_{xz} + \chi^{(0)}_{yz})] 
\nonumber\\
&-& v_q^2 e^{-2 q d}\chi^{(0)}_{xy}(\chi^{(0)}_{xz} + \chi^{(0)}_{yz})~,
\end{eqnarray}
and for brevity we have suppressed the $({\bm q},\omega) $ dependence of the three non-interacting density-response functions appearing in Eq.~(\ref{eq:RPAdielectric}). 

In Figure~\ref{fig:Feynman} we present the Feynman diagrams which contribute to the $G_0W$-RPA self-energy of a quasiparticle in one of the elliptical bands of the $t_{2g}$ 2DEG. Summing this infinite series of bubble diagrams (with all of the directed lines representing the non-interacting Green's functions omitted) offers a second route to deriving the RPA interaction $W_{xz}({\bm q},\omega)$. From Figure~\ref{fig:Feynman} we see that density fluctuations in both the circular $xy$ band and the elliptical $yz$ and $xz$ bands contribute to screening, and that density fluctuations in bands with different (the same) $\hat{\bm z}$-direction confinement, interact with each other via $v_qe^{-q d} $ ($v_q $). When the Green's functions are included in Figure~\ref{fig:Feynman}, the series sums to the full $G_0W$-RPA self-energy. The diagrammatic representation emphasizes that $G_0W$ can be viewed~\cite{hedin_pr_1965} as an expansion of the self-energy to lowest order in the \emph{screened} electron-electron interaction $W_{xz}({\bm q},\omega)$.    

The finite-temperature~\cite{Fetter_and_Walecka} $G_0W$-RPA self-energy of a quasiparticle in band $\alpha$ is given by
\begin{eqnarray}\label{eq:Matsubara_SE}
\Sigma_{\alpha}({\bm k},i\omega_n) &=& -\frac{1}{\beta \hbar A}\sum_{{\bm q},i\Omega_m}W_{\alpha}({\bm q}, i\Omega_m) \nonumber\\
&\times & G^{(0)}_{\alpha}({\bm k} - {\bm q}, i\omega_n - i\Omega_m)~,
\end{eqnarray}
where $\beta = (k_{\rm B} T)^{-1}$ and the fermionic $\omega_n$ and bosonic $\Omega_m$ Matsubara frequencies are given by $\omega_n = (2n+1)\pi/\hbar\beta$ and $\Omega_m = 2m\pi/\hbar \beta$, respectively. The non-interacting Green's function is given by 
\begin{equation}\label{eq:Greensfunction}
G^{(0)}_{\alpha}({\bm k},i\omega_n) = \frac{1}{i\omega_n - \xi_{\alpha}({\bm k})/\hbar}~.
\end{equation}

The physical properties of the interacting system depend on the {\it retarded} self-energy, which can be obtained from Eq.~(\ref{eq:Matsubara_SE}) via analytic continuation, $i\omega_n \rightarrow \omega + i\eta $, only \emph{after} carrying out the Matsubara frequency summation over $\Omega_m$. The so-called line-residue decomposition~\cite{Quinn_and_Ferrell} 
proceeds in the 
reverse order. By carrying out the analytic continuation of Eq.~(\ref{eq:Matsubara_SE}) {\it before} 
the frequency summation we obtain the (purely real) {\it line} contribution to the retarded self-energy
\begin{eqnarray}\label{eq:line_term}
\Sigma^{\rm line}_{\alpha}({\bm k},\omega) &=& - \frac{1}{(2 \pi)^3}\int_{-\infty}^{\infty}d\Omega \int d^2{\bm q}~W_{\alpha}({\bm q}, i\Omega) \nonumber\\
&\times & G^{(0)}_{\alpha}({\bm k}-{\bm q},\omega - i\Omega)~.
\end{eqnarray}
The {\it residue} contribution corrects for performing the analytic continuation before evaluating the frequency sum, and is given by 
\begin{eqnarray}\label{eq:freqsum}
\Sigma^{\rm res}_{\alpha}({\bm k},\omega)&=& -\frac{1}{ (2 \pi)^2}\int d^2{\bm q}~W_{\alpha}({\bm q},\omega - 
\xi_{\alpha}({\bm k}-{\bm q})/\hbar) \nonumber\\
&\times& \left[ \Theta \left\{-\xi_{\alpha}({\bm k}-{\bm q}) \right\} - \Theta \left\{\hbar\omega-\xi_{\alpha}({\bm k}-{\bm q})\right\}\right]~,\nonumber\\
\end{eqnarray}
where $\Theta \left\{x \right\}$ is the Heaviside step function. In the next two sections we calculate two important properties of $t_{2g}$ 2DEG quasiparticles based on this formulation of the $G_0W$-RPA retarded self-energy.

\section{Fermi Surface Shape Modification (FSSM)}
\label{Sect:FSrenorm}
Consider an ordinary single-band isotropic 2DEG~\cite{Giuliani_and_Vignale}. 
When interactions are adiabatically turned on,  
all quasiparticles on the non-interacting Fermi surface have infinite lifetime and experience identical
energy shifts given by Eq.~(\ref{eq:QP_EE}) and equal to the interaction contribution to the chemical potential.
Because of rotational invariance, the self-energy contribution is a function of the magnitude of ${\bm k}$ only.
All isoenergy surfaces, including the Fermi surface, continue to be circular in the interacting system. 
Furthermore, Luttinger's theorem~\cite{Luttinger} constrains the Fermi surface area of interacting quasiparticles to equal the Fermi surface area of non-interacting electrons, leading to the conclusion that interactions do not 
yield a Fermi surface shape modification (FSSM).  This simplification is artificial however, since interacting electron
systems in solids are never perfectly isotropic.
In the presence of anisotropy, the self-energy contribution to the quasiparticle energy spectrum is dependent on the orientation of ${\bm k} $, and the Fermi surface shape can therefore be renormalized by interactions. 

We expect this phenomena to be relevant for the anisotropic $xz$ and $yz$ bands of the $t_{2g}$ electron gas. 
Each band has an elliptical non-interacting Fermi surface. 
In the following calculations we assume that the renormalized Fermi surface is
sufficiently close in shape to an ellipse, that it can still be characterized by 
two wavevectors, $k^{\ast}_{{\rm F}x}$ and $k^{\ast}_{{\rm F} y}$, whose values are renormalized by interactions 
from their non-interacting values $k_{{\rm F} x}$ and $k_{{\rm F} y}$. 
Below we explicitly discuss FSSM for the 
$xz$ band, which has its semimajor axis in the $\hat{\bm y}$-direction and semiminor axis in the $\hat{\bm x}$-direction. Results for the $yz$ band can be found by interchanging $k^{\ast}_{{\rm F} x}$ and $k^{\ast}_{{\rm F} y}$.

Our main finding is that Fermi surface anisotropy is reduced by interactions (see Figs.~\ref{fig:two} and~\ref{fig:four}). This result can be understood qualitatively at the Hartree-Fock level. 
The exchange (X) self-energy of the $xz$-band  is given by
\begin{equation} \label{eq:Sigma_ex}
\Sigma^{\rm X}_{xz}({\bm k})= - \frac{1}{(2 \pi)^2}\int d^2{\bm q} \frac{2 \pi e^2}{\kappa \vert{\bm k}-{\bm q} \vert}\Theta\left\{\varepsilon_{{\rm F}xz} - \varepsilon_{xz}(\bm q)\right\}~,
\end{equation} 
where $\varepsilon_{{\rm F}xz}$ is the Fermi energy. Eq.~(\ref{eq:Sigma_ex}) can be easily obtained from Eq.~(\ref{eq:Matsubara_SE}) by replacing the dynamically-screened interaction $W_{xz}({\bm q}, i\Omega_m)$ with the bare Coulomb interaction $v_q$.
From Eq.~(\ref{eq:Sigma_ex}) we see that a quasiparticle with quantum number ${\bm k}$ will have a
self-energy correction that is larger in magnitude
when there are more occupied states nearby in momentum space;
because of the Coulomb interaction factor, the integrand is large for ${\bm q}$ near ${\bm k}$, but only if the 
state ${\bm q}$ is occupied. 
Since the $xz$ band's non-interacting Fermi surface is elliptical, with its semimajor axis parallel to the $\hat{\bm y}$-axis, an electron at the Fermi surface in the $\hat{\bm y}$-direction will have fewer occupied states in its neighborhood than an electron at the Fermi surface in the $\hat{\bm x}$-direction.  It follows that 
\begin{equation}
\Sigma^{\rm X}_{xz}(k_{{\rm F}x}) < \Sigma^{\rm X}_{xz}(k_{{\rm F}y})~,
\end{equation}
where we note that the exchange self-energy is always negative. 
Since all states ${\bm k}$ sitting on the Fermi surface must, by definition, have the same quasiparticle energy,
the Fermi surface will change shape when interactions are taken into account and the degree of Fermi surface anisotropy 
will be reduced. Below we report numerical calculations which include beyond Hartree-Fock contributions to the 
quasiparticle self-energy that confirm this expectation. 

\subsection{Linearized self-energy estimate of FSSM}
\label{Sect:Linearized}
We begin with a numerical approach valid for 
weak interactions that is rather simple to implement.
In Section~\ref{Sect:Self_Consistent} we solve the problem self-consistently. 
We will find the two methods give nearly identical results.

\begin{figure}[t]
\includegraphics[width=1.0\linewidth]{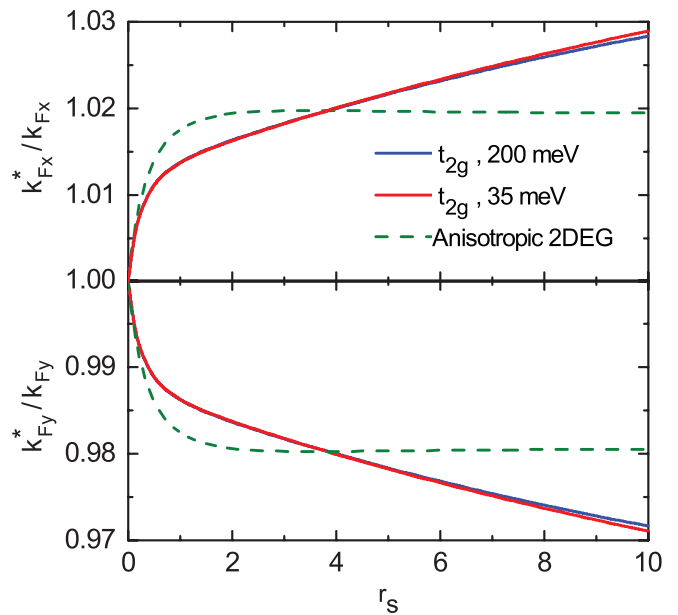}
\caption{(Color online) Fractional change in the $xz$ band's Fermi wavevectors, $k^{\ast}_{{\rm F} x}/k_{{\rm F}x}$ (top) and $k^{\ast}_{{\rm F}y}/k_{{\rm F}y}$ (bottom) as a function of interaction strength parameter $r_{\rm s}$. 
The red curve is for band offset $\Delta = 35~{\rm meV}$ and layer separation $d=10a$ where $a=3.9$~\AA is the SrTiO$_3$ lattice constant.  The blue curve is for band offset $\Delta = 200~{\rm meV}$ 
and layer separation $d=2a$. The dashed green curve is for a single-band anisotropic 2DEG whose $\hat{\bm x}$-direction and $\hat{\bm y}$-direction non-interacting masses are the 
same as those of the $xz$ band in the $t_{2g}$ 2DEG model. 
For this curve $r_{\rm s}$ is defined from the total density in the single anisotropic band. \label{fig:two}}
\end{figure}

When the change in Fermi surface is small relative to its original dimensions, we are well
justified in expanding Eq.~(\ref{eq:chem_pot}) to linear-order in 
$\delta{\bm k}_{\rm F}  \equiv {\bm k}^{\ast}_{\rm F} - {\bm k}_{\rm F}$. 
In the limit $\delta k_{{\rm F}x}/k_{{\rm F}x} \ll 1$ we have 
\begin{equation}\label{eq:Linone}
\begin{array}{l}
{\displaystyle \mu = \varepsilon_{xz}(k_{{\rm F} x}) + \Sigma_{xz}(k_{{\rm F}x}, 0)}\vspace{0.2 cm} \\
{\displaystyle \qquad \qquad + \; \delta k_{{\rm F}x}\partial_{k_x}\left[\varepsilon_{xz}({\bm k}) + \Sigma_{xz}({\bm k},0) \right]_{{\bm k} = k_{{\rm F} x}}}~.
\end{array}
\end{equation}
Similarly for  $\delta k_{{\rm F}y}/k_{{\rm F} y} \ll 1$ we have
\begin{equation}\label{eq:Lintwo}
\begin{array}{l}
{\displaystyle \mu = \varepsilon_{xz}(k_{{\rm F} y}) + \Sigma_{xz}(k_{{\rm F} y},0)}\vspace{0.2 cm}\\ 
{\displaystyle \qquad \qquad + \; \delta k_{{\rm F}y} \, \partial_{k_y}\left[\varepsilon_{xz}({\bm k}) + \Sigma_{xz}({\bm k},0) \right]_{{\bm k} = k_{{\rm F} y}}~.}
\end{array}
\end{equation}
%
For fixed total elliptical band density 
\begin{equation}\label{eq:Linthree}
k_{{\rm F} y}\delta k_{{\rm F}x} + k_{{\rm F} x}\delta k_{{\rm F}y} = 0~.
\end{equation}
The three previous equations can be solved for $\mu$, $\delta k_{{\rm F}x}$, and $\delta k_{{\rm F}y}$
given self-energy values and wavevector derivatives on the non-interacting Fermi surface. In the top and bottom panel of Fig.~\ref{fig:two} we plot 
$k^{\ast}_{{\rm F} x}/k_{{\rm F}x}$ and $k^{\ast}_{{\rm F}y}/k_{{\rm F}y}$ versus an effective
interaction strength parameter $r_{\rm s}$ defined following
the convention commonly used in the single-band 2DEG 
literature~\cite{Giuliani_and_Vignale}:
\begin{equation}\label{eq:definers}
n_{xz} = \frac{1}{\pi (a^*_{\rm B} r_{\rm s})^2}
\end{equation}
where $a^*_{\rm B} = \kappa \, a_{\rm B}/m_{\rm DOS}$ defines the effective ({\it i.e.} material) Bohr radius,
$a_{\rm B}$ is the atomic Bohr radius, and $m_{\rm DOS}=\sqrt{m_{\rm H} m_{\rm L}}$ has been introduced earlier. As expected, we find that interactions tend to reduce the anisotropy of the elliptical bands in the $t_{2g}$ 2DEG. For comparison, we also plot in Fig.~\ref{fig:two} renormalized Fermi wavevectors calculated for a single-band anisotropic 2DEG with energy dispersion given by
\begin{equation}
\varepsilon({\bm k}) = \frac{\hbar^2 k^2_x}{2 m_{\rm L}} + \frac{\hbar^2 k^2_y}{2 m_{\rm H}}
\end{equation}
and density $n=n_{xz}$.  

When $r_{\rm s}$ is small, the amount of FSSM is similar in both the $xz$ band of the $t_{2g}$ 2DEG and in the single-band anisotropic 2DEG. This occurs for two reasons. First, $\partial_{k_x}\left. \varepsilon_{xz}({\bm k})\right|_{{\bm k} = k_{{\rm F} x}} $ and $\partial_{k_y}\left. \varepsilon_{xz}({\bm k})\right|_{{\bm k} = k_{{\rm F} y }} $ are dominant over $\partial_{k_x}\left. \Sigma_{xz}({\bm k},0) \right|_{{\bm k} = k_{{\rm F} x}} $ and $\partial_{k_y}\left. \Sigma_{xz}({\bm k},0) \right|_{{\bm k} = k_{{\rm F} y }} $, respectively, in Eq.~(\ref{eq:Linone}) and Eq.~(\ref{eq:Lintwo}) when  $r_{\rm s} \ll 1$. And second, because the leading-order contribution to the self-energy when $r_{\rm s} \ll 1$ is the exchange self-energy of Eq.~(\ref{eq:Sigma_ex}), which is equal for both the $xz$ band of the $t_{2g}$ 2DEG and the single-band anisotropic 2DEG. The small $r_{\rm s}$ limit allows for simple explanation because both systems are weakly interacting and well described at leading-order by Hartree-Fock.

The situation is more interesting at large values of $r_{\rm s}$ where correlation effects are important. Specifically, Figure~\ref{fig:two} suggests that while FSSM in the single-band 2DEG saturates at large $r_{\rm s}$, FSSM in the anisotropic bands of the $t_{2g}$ 2DEG increases with increasing $r_{\rm s}$. As we discuss in detail in the next section, the difference in FSSM occuring in $t_{2g}$ system and in the ordinary 2DEG depends sensitively on the influence of the additional screening due to the presence of several occupied bands. 
\subsection{FSSM at low electron density}

Given the $G_0W$-RPA description of quasiparticles, the principle difference between the anisotropic 2DEG and the $xz$ band of the $t_{2g}$ 2DEG, is that the latter also has electrons occupying the $yz$ and $xy$ bands. To understand how these other occupied conduction bands produce additional screening, and how this screening then qualitatively changes FSSM when density is low, in this section we derive analytic expressions for the self-energy and the wavevector derivative of the self-energy evaluated on the Fermi surface, to leading-order in the small-parameter $1/r_{\rm s} $. These expressions reveal the basic physical mechanisms which govern FSSM at large $r_{\rm s}$ within $G_0W$-RPA theory.

Many of the formulas in this section are written out explicitly for the $xz$ band of the $t_{2g}$ 2DEG, however, analogous expressions for the single-band anisotropic 2DEG can be found by replacing the self-energy of the $xz$ band with the self-energy of the single-band 2DEG. In our numerical calculations, this is carried out simply by setting to zero the density in the $xy$ and $yz$ bands.

 We begin by approximating Eq.~(\ref{eq:Linone}) and Eq.~(\ref{eq:Lintwo}) as 
\begin{equation}\label{eq:deltakx}
\frac{\delta k_{{\rm F}x}}{k_{{\rm F}x}} = \frac{1}{2}\frac{\Sigma_{xz}(k_{{\rm F}y},0)-\Sigma_{xz}(k_{{\rm F}x},0)}{k_{{\rm F}}\left.\partial_{k}\Sigma_{xz}(k,0) \right|_{k = k_{\rm F}} }
\end{equation}
and
\begin{equation}\label{eq:deltaky}
\frac{\delta k_{{\rm F}y}}{k_{{\rm F}y}} = \frac{1}{2}\frac{\Sigma_{xz}(k_{{\rm F}x},0)-\Sigma_{xz}(k_{{\rm F}y},0)}{k_{{\rm F}} \left.\partial_{k}\Sigma_{xz}(k,0) \right|_{k = k_{\rm F}} }~.
\end{equation}
 In obtaining Eq.~(\ref{eq:deltakx}) we made the approximations
\begin{equation}\label{eq:numerone}
\mu - \varepsilon_{xz}(k_{{\rm F}x}) - \Sigma_{xz}(k_{{\rm F}x},0) \to \frac{\Sigma_{xz}(k_{{\rm F} y},0)-\Sigma_{xz}(k_{{\rm F}x},0)}{2} 
\end{equation}
and
\begin{equation}\label{eq:numertwo}
\partial_{k_x}\left[ \varepsilon_{xz}({\bm k}) + \Sigma_{xz}({\bm k},0) \right]_{{\bm k} = k_{{\rm F}x}} \rightarrow \left.\partial_{k}\Sigma_{xz}(k,0) \right|_{ k = k_{\rm F}}
\end{equation}
where $\partial_{k}\Sigma_{xz}(k,0)$ is calculated using an alternative version of the $t_{2g}$ 2DEG model \emph{without anisotropy}. Specifically, for this particular quantity we use the isotropic version of Eq.~(\ref{eq:ellipticalbands}):
\begin{equation}\label{eq:ellipticalbands_two}
\varepsilon_{\alpha}({\bm k}) = \left\{
\begin{array}{l}
{\displaystyle   \frac{\hbar^2 {\bm k}^2}{2 m_{\rm L}}\qquad   \qquad \; \; \; \; {\rm for} \qquad \alpha=xy,\sigma}\vspace{0.1 cm}\\
{\displaystyle   \frac{\hbar^2 {\bm k}^2}{2 m_{\rm DOS}} + \Delta \qquad  {\rm for} \qquad \alpha=xz,\sigma}\vspace{0.1 cm}\\
{\displaystyle \frac{\hbar^2 {\bm k}^2}{2 m_{\rm DOS}} + \Delta \qquad {\rm for} \qquad \alpha=yz,\sigma}
\end{array}
\right.
\end{equation}
The $xz$ bands Fermi wavevector in this isotropic version is $k_{\rm F} \equiv \sqrt{k_{{\rm F}x}k_{{\rm F}y}} $. Analogous approximations were carried out to obtain Eq.~(\ref{eq:deltaky}), and we have confirmed numerically that the large $r_{\rm s} $ asympotic behavior appearing in Figure~\ref{fig:two} is qualitatively unaltered by these approximations. 

In a moment we will explain the differences in FSSM between the single-band anisotropic 2DEG and the $xz$ band of the $t_{2g}$ 2DEG by separately calculating the numerator and denominator of Eqs.~(\ref{eq:deltakx}) and~(\ref{eq:deltaky}) to leading-order in powers of $1/r_{\rm s}$. First let us consider a simple scaling argument which highlights the critical role played by non-analyticity in the $G_0W$-RPA self-energy of the $t_{2g}$ 2DEG. Consider the following series representation for the self-energy at zero frequency:
\begin{equation}\label{eq:Sigma_rs_series}
\Sigma_{xz}({\bm k},0) = \frac{f({\bm k})}{r^{\alpha}_{\rm s}} + \frac{g({\bm k})}{r^{\beta}_{\rm s}} + {\cal O}\left( \frac{1}{r^{\gamma}_{\rm s}} \right)  + \ldots
\end{equation}
where $\alpha < \beta < \gamma $. We expect that the leading-order contribution to the numerator of Eqs.~(\ref{eq:deltakx}) and~(\ref{eq:deltaky}) comes from the leading-order term in Eq.~(\ref{eq:Sigma_rs_series}) with a coefficient ({\it e.g.} $f({\bm k})$ or $g({\bm k})$) that actually depends on ${\bm k}$, as opposed to being a constant. Our calculations reveal that $\partial_kf({\bm k}) = 0$ for both the $xz$ band of the $t_{2g}$ 2DEG and the single-band anisotropic 2DEG, and therefore the leading-order contribution to the numerator of Eqs.~(\ref{eq:deltakx}) and~(\ref{eq:deltaky}) comes from the sub-leading term in Eq.~(\ref{eq:Sigma_rs_series}):
\begin{equation}\label{eq:scaling_one}
\Sigma_{xz}(k_{{\rm F} y},0)-\Sigma_{xz}(k_{{\rm F}x},0) \sim \frac{g(k_{{\rm F}y}) - g(k_{{\rm F}x})}{r^{\beta}_{\rm s}}~.
\end{equation}
If the $G_0W$-RPA self-energy ({\it i.e.}~$g({\bm k})$) is analytic for ${\bm k}$ on the non-interacting Fermi surface, then we can obtain the leading-order term in the $1/r_{\rm s}$ series expansion for the denominators of Eqs.~(\ref{eq:deltakx}) and~(\ref{eq:deltaky}) directly from Eq.~(\ref{eq:Sigma_rs_series}). This gives
\begin{equation}\label{eq:scaling_two}
k_{{\rm F}}\left.\partial_{k}\Sigma_{xz}(k,0) \right|_{k = k_{\rm F}} \sim  \frac{g(k_{\rm F})}{r^{\beta}_{\rm s}}~,
\end{equation}
where we have approximated the wavevector derivative of the self-energy by dividing through by the Fermi wavevector $k_{\rm F} $. When Eqs.~(\ref{eq:scaling_one}) and~(\ref{eq:scaling_two}) are substituted into Eqs.~(\ref{eq:deltakx}) and~(\ref{eq:deltaky}) we obtain a constant,
\begin{equation}\label{eq:deltakx_scaling}
\frac{\delta k_{{\rm F}x}}{k_{{\rm F}x}} = -\frac{\delta k_{{\rm F}y}}{k_{{\rm F}y}} = \frac{g(k_{{\rm F}y}) - g(k_{{\rm F}x})}{2 g(k_{\rm F})}~,
\end{equation}
which while describing perfectly the saturation of FSSM in the single-band anisotropic 2DEG at large $r_{\rm s}$, fails to describe the $t_{2g}$ 2DEG. As we show explicitly below, the leading-order contribution to the wavevector dependent part of the $G_0W$-RPA self-energy of the $xz$ band of the $t_{2g}$ 2DEG is not analytic at the Fermi surface, and therefore the simple arguments leading to Eq.~(\ref{eq:deltakx_scaling}) do not apply in this case. The origin of this divergence is the long-range of the Coulomb interaction, and is similar to the divergence of the quasiparticle effective mass within Hartree-Fock theory~\cite{Giuliani_and_Vignale}. 

We begin by calculating the denominator of Eqs.~(\ref{eq:deltakx}) and~(\ref{eq:deltaky}) for the single-band 2DEG. Putting wavevectors and frequencies in units of $k_{\rm F}$ and $\hbar \, k^2_{\rm F}/m_{\rm DOS} $, respectively, we obtain
\begin{eqnarray}\label{eq:ddkSigma_2DEG_one}
\left.\partial_{k}\Sigma(k,0) \right|_{k = k_{\rm F}}  = \left(\frac{2 a_{\rm B} {\rm Ry}}{\pi^2 \kappa}\right) \int^{\infty}_{0} \! q dq \int^{\pi}_{-\pi} \! d\theta \int^{\infty}_{0} \! d\Omega \nonumber\\
 \times \frac{1/q}{1+\frac{2^{1/2} r_{\rm s}}{q} \chi^{(0)}(q,i\Omega)} 
 \left(\frac{2 q \cos{\left[ \theta \right]} -2 }{\left\{q^2 - 2 q \cos{\left[\theta\right]} + 2i\Omega\right\}^2}\right),
\end{eqnarray}
where we have removed the $xz$ subscript from the self-energy to explicitly indicate we are here considering the single-band 2DEG. The dimensionless Lindhard function $\chi^{(0)}(q,i\Omega)$ is obtained from the dimensionful form~\cite{Giuliani_and_Vignale} by dividing out the negative density-of-states. We must now consider how to extract the leading-order in powers of $1/r_{\rm s}$ term in Eq.~(\ref{eq:ddkSigma_2DEG_one}).  Physically speaking, a quasiparticle at momentum ${\hbar\bm k}$ and energy $\hbar\omega$ acquires its self-energy by making virtual transitions to intermediate states of momentum ${\hbar \bm k}-{\hbar \bm q}$ and energy $\hbar\omega-\hbar\Omega$, and back. This picture is motivated by the Feynman diagram for the $G_0W$-RPA self-energy and its mathematical representation, Eq.~(\ref{eq:Matsubara_SE}). The available phase-space for virtual transitions in which the momentum and energy transferred is on the order of the Fermi momentum and Fermi energy, respectively, scales like $k_{\rm F}^4 \propto 1/r^4_{\rm s}$, and vanishes swiftly in the limit of low density. Therefore the leading-order in powers of $1/r_{\rm s}$ term from Eq.~(\ref{eq:ddkSigma_2DEG_one}) instead comes from $\vert {\bm q} \vert \gg 1 $ and $\Omega \gg 1 $. In this limit the dimensionless Lindhard function has a particularly simple form~\cite{Giuliani_and_Vignale}
\begin{equation}
\chi^{(0)}(q,i\Omega) \rightarrow \frac{2 q^2}{q^4 + 4 \Omega^2}~.
\end{equation}
After expanding Eq.~(\ref{eq:ddkSigma_2DEG_one}) to leading-order in the small parameter $\cos{(\theta)}/q$ we obtain an expression which can be evaluated analytically. We then find
\begin{equation}\label{eq:ddkSigma_2DEG_two}
\left.\partial_{k}\Sigma(k,0) \right|_{k = k_{\rm F}}  = \left(\sqrt{\frac{2}{\pi}}\frac{\Gamma\left[\frac{1}{3}\right]\Gamma\left[\frac{7}{6}\right]a_{\rm B}{\rm Ry}}{\kappa}\right)\frac{1}{r^{1/3}_{\rm s}}
\end{equation}
where $\Gamma\left[x\right]$ is the Euler gamma function. Next we evaluate the numerator of Eqs.~(\ref{eq:deltakx}) and~(\ref{eq:deltaky}) for the single-band anisotropic 2DEG. Following similar steps, and expanding the self-energy to leading-order in $\zeta \equiv m_{\rm H}/m_{\rm L}$ about $\zeta=1$ we obtain
\begin{eqnarray}\label{eq:ddkSigma_2DEG_three}
\Sigma(k_{{\rm F} y},0)-\Sigma(k_{{\rm F}x},0)&=& -\left(\frac{8\,\Gamma\left[\frac{7}{6}\right]\Gamma\left[-\frac{8}{3}\right] \bar{m}_{\rm DOS}{\rm Ry}}{81 \sqrt{\pi}  \kappa^2}\right) \nonumber\\
&\times & \frac{\zeta-1}{r^{4/3}_{\rm s}}~,
\end{eqnarray}
where $\bar{m}_{\rm DOS}=m_{\rm DOS}/m$ and $m$ is the bare electron mass in vacuum. We have successfully compared both Eqs.~(\ref{eq:ddkSigma_2DEG_two}) and~(\ref{eq:ddkSigma_2DEG_three}) against the full $G_0W$-RPA numerical calculations, and when these expressions are substituted into Eqs.~(\ref{eq:deltakx}) and~(\ref{eq:deltaky}) we confirm that FSSM in the single-band anisotropic 2DEG saturates at large $r_{\rm s}$.

Let us now move on to considering the leading-order in $1/r_{\rm s}$ expressions for the numerator and denominator of Eqs.~(\ref{eq:deltakx}) and~(\ref{eq:deltaky}) for the $xz$ band of the $t_{2g}$ 2DEG. Once again, phase-space considerations for virtual transitions suggest that the most important ${\hbar \bm q}$ and $\hbar \Omega$ are large on the scale of the $xz$ band's Fermi momentum and Fermi energy, respectively. As a result, the RPA screened Coulomb interaction is accurately approximated by a simplified version of the dielectric function given in Eq.~(\ref{eq:RPAdielectric}) which neglects $\chi^{(0)}_{xz}(q,i\Omega)$ and $\chi^{(0)}_{yz}(q,i\Omega)$ compared to $\chi^{(0)}_{xy}(q,i\Omega)$. It is useful to define the dimensionless interaction parameter $R_{\rm s} $, which is analogous to Eq.~(\ref{eq:definers}) but describes the density in the $xy$ band 
\begin{equation}\label{eq:defineRS}
n_{xy} = \frac{1}{\pi (a^*_{{\rm B}xy} R_{\rm s})^2}~.
\end{equation}
Here $a^*_{{\rm B}xy} = \kappa \, a_{\rm B}/m_{\rm L}$ defines the effective Bohr radius appropriate for the $xy$ band. In the limit of a large energy offset, $\Delta$, between the bottom of the $xz$ and $yz$ bands and the bottom of the $xy$ band ({\it i.e.}~$R_{\rm s} \ll 1 $), we are justified in approximating $\chi^{(0)}_{xy}(q,i\Omega)$ by its long wavelength limit. As detailed in the Appendix, we then find
\begin{eqnarray}\label{eq:ddkSigma_2DEG_four}
\left.\partial_{k}\Sigma_{xz}(k,0) \right|_{k = k_{\rm F}}  &=& -\left(\frac{a_{\rm B} m_{\rm L} {\rm Ry}}{2  \pi \, \kappa \,  m_{\rm DOS}}\right)\nonumber \\
&\times&\frac{R_{\rm s}}{r_{\rm s}}\ln{\left[\frac{\kappa^2}{2 \bar{m}_{\rm L}} \Delta + \frac{m_{\rm DOS}}{m_{\rm L}}\frac{1}{r^2_{\rm s}} \right]}
\end{eqnarray}
where $\Delta$ is in units of Rydbergs and $\bar{m}_{\rm L} = m_{\rm L}/m $. Similar approximations allow us to evaluate the numerator of Eqs.~(\ref{eq:deltakx}) and~(\ref{eq:deltaky}):
\begin{equation}\label{eq:ddkSigma_2DEG_five}
\Sigma_{xz}(k_{{\rm F} y},0)-\Sigma_{xz}(k_{{\rm F}x},0) = \left(\frac{2^{3/2} \bar{m}_{\rm DOS} {\rm Ry}}{\pi^2 \kappa^2}\right){\cal F}(\zeta) \frac{R_{\rm s}}{r^2_{\rm s}}
\end{equation}
where the function ${\cal F}(\zeta)$ can be written in terms of complete elliptic integrals and is given explicitly in the Appendix. After substituting Eqs.~(\ref{eq:ddkSigma_2DEG_four}) and~(\ref{eq:ddkSigma_2DEG_five}) into Eqs.~(\ref{eq:deltakx}) and~(\ref{eq:deltaky}) we find that FSSM in the elliptical bands of the $t_{2g}$ 2DEG grows (sublinearly) with increasing $r_{\rm s}$ until the density in these bands is so low that $m_{\rm DOS}/ (m_{\rm L} r^2_{\rm s}) \ll  \kappa^2\Delta/(2 \bar{m}_{\rm L} )$, at which point FSSM in the elliptic bands of the $t_{2g}$ 2DEG also saturates. Numerical calculations show that FSSM of the elliptic bands does indeed saturate for values of $r_{\rm s}$ much greater than those shown in Figure~\ref{fig:two}. The presence of the logarithmic factor in Eq.~(\ref{eq:ddkSigma_2DEG_four}) is a signature of the non-analyticity of the $xz$ band's $G_0W$-RPA self-energy when $R_{\rm s} \ll 1 $, and gives a simple explanation for FSSM in the elliptical bands of the $t_{2g}$ 2DEG when $r_{\rm s}$ is large. 
\subsection{Self-consistent calculation of FSSM}
\label{Sect:Self_Consistent}
\begin{figure}[h]
\includegraphics[width=1.0\linewidth]{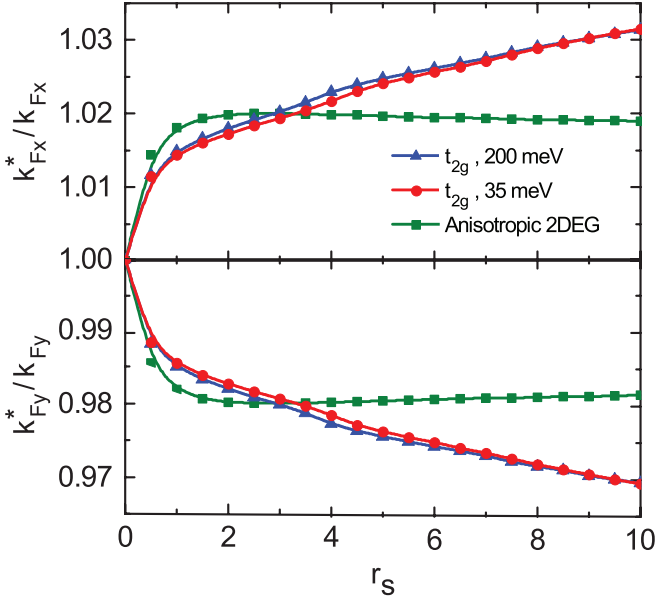}
\caption{(Color online) Same as Fig.~\ref{fig:two}, but calculated by the self-consistent solution of Eqs.~(\ref{eq:SCone}) and~(\ref{eq:SCtwo}), rather than using the linearized Eqs.~(\ref{eq:deltakx})-(\ref{eq:deltaky}). \label{fig:four}}
\end{figure}

Finally, to confirm the FSSM results in Section~\ref{Sect:Linearized}, and also to check the validity of assuming that FSSM is small, we solve for the $xz$ band's renormalized Fermi surface by 
self-consistently solving the two equations
\begin{equation}\label{eq:SCone}
\mu=\varepsilon_{xz}(k^{\ast}_{{\rm F} x}) + {\rm Re}\Sigma_{xz}(k^{\ast}_{{\rm F} x},0)
\end{equation}
and
\begin{equation}\label{eq:SCtwo}
\mu=\varepsilon_{xz}(k^{\ast}_{{\rm F} y}) + {\rm Re}\Sigma_{xz}(k^{\ast}_{{\rm F} y},0)~.
\end{equation}
The numerical solution follows simply by iteratively solving the above two equations while forcing the ${\bm k}$-space area of the $xz$ band to remain constant. In the top and bottom panels of Fig.~\ref{fig:four} we plot $k^{\ast}_{{\rm F} x}/k_{{\rm F}x}$ and $k^{\ast}_{{\rm F} y}/k_{{\rm F}y}$ versus $r_{\rm s}$. Clearly the conclusions of Section~\ref{Sect:Linearized} are confirmed. 

We now turn our attention to understanding how interactions effect the quasiparticle effective mass. 
\section{Quasiparticle Effective Mass}
\label{Sect:Mass}

Just as the free-space electron mass is renormalized by the periodic crystal potential of a solid, it can further be 
renormalized by the presence of electron-electron interactions~\cite{Pines_and_Nozieres,Giuliani_and_Vignale}. In both cases, the renormalized mass is defined so that the single-particle excitation energies are well approximated by the non-interacting kinetic energy formula with the bare mass replaced by the renormalized mass. In the $t_{2g}$ 2DEG, the effect of the crystal potential is already captured by Eq.~(\ref{eq:ellipticalbands}). To evaluate the effect of electron-electron interactions on the $t_{2g}$ 2DEG quasiparticle mass values, we expand the quasiparticle excitation energy about the renormalized Fermi surface:
\begin{equation}\label{eq:Taylor}
\mathcal{E}_{\alpha}({\bm k}) = ({\bm k}-{\bm k}^{\ast}_{{\rm F}\alpha}) \cdot \nabla_{\bm k}\mathcal{E}_{\alpha}({\bm k})\vert_{{\bm k} = {\bm k}^{\ast}_{{\rm F}\alpha}}  + \mathcal{O}(\delta{\bm k}^2)~.
\end{equation}
We focus on the quasiparticle mass of the $xz$ band for which the $\hat{\bm x}$-direction bare mass is $m_{\rm L}$ (including the effect of the periodic crystal potential but not yet including electron-electron interactions) and the $\hat{\bm y}$-direction bare mass is $m_{\rm H}$. 

The renormalized light and heavy masses can be calculated using 
\begin{equation}\label{eq:MassL}
m^{\ast}_{\rm L} = \frac{\hbar^2 k_{{\rm F}x}^{\ast}}{\left[d\mathcal{E}_{xz}({\bm k})/dk_x\right]_{{\bm k}=k_{{\rm F}x}^{\ast}}}
\end{equation}
and
\begin{equation}\label{eq:MassH}
m^{\ast}_{\rm H} = \frac{\hbar^2 k_{{\rm F}y}^{\ast}}{\left[d \mathcal{E}_{xz}({\bm k})/dk_y\right]_{{\bm k}=k_{{\rm F}y}^{\ast}}}~.
\end{equation}
Using Eq.~(\ref{eq:QP_EE}) relating the quasiparticle energy to the self-energy, we find the following relationship between the self-energy and the quasiparticle masses,
\begin{equation}\label{eq:mLeq}
\frac{m^{\ast}_{\rm L}}{m_{\rm L}} = \frac{1- \hbar^{-1}\partial_{\omega}{\rm Re}\Sigma_{xz}(k_{{\rm F}x}^{\ast},\omega)\vert_{\omega = 0}}{1 + \left(m_{\rm L}/\hbar^2k^{\ast}_{{\rm F}x}\right)\partial_{k_x}{\rm Re}\Sigma_{xz}({\bm k},0) \vert_{{\bm k}=k_{{\rm F}x}^{\ast}} }
\end{equation}
and
\begin{equation}\label{eq:mHeq}
\frac{m^{\ast}_{\rm H}}{m_{\rm H}} = \frac{1- \hbar^{-1}\partial_{\omega}{\rm Re}\Sigma_{xz}(k_{{\rm F}y}^{\ast},\omega)\vert_{\omega = 0}}{1 + \left(m_{\rm H}/\hbar^2k^{\ast}_{{\rm F}y}\right)\partial_{k_y}{\rm Re}\Sigma_{xz}({\bm k},0)\vert_{{\bm k}=k_{{\rm F} y}^{\ast}}}~.
\end{equation}
The right-hand-sides of these equations can be evaluated using the $G_0W$-RPA approximation for the self-energy.

\subsection{Quasiparticle effective mass at high densities}
\label{Sect:Analytic}
Before presenting our numerical results, we briefly consider the high-density (small $r_{\rm s}$) limit where analytic results for the quasiparticle mass can be obtained. 
These results will offer insight into the physical processes which renormalize the electron mass.

We start by simplifying the $t_{2g}$ model slightly to make the derivation tractable. We take the $xz$ and $yz$ bands to be isotropic with mass $m_{\rm DOS} = \sqrt{m_{\rm H}m_{\rm L}}$ as in Eq.~(\ref{eq:ellipticalbands_two}). The influence of anisotropy on the quasiparticle mass is isolated and studied in Section~\ref{Sect:Anisotropy}, but we neglect it here. In this section it is advantageous to work with the zero-temperature formalism of many-body perturbation theory~\cite{Fetter_and_Walecka}. In units of effective Rydbergs, ${\rm Ry}^{\ast} = (m_{\rm DOS}/\kappa^2){\rm Ry}$, the $G_0W$-RPA self-energy of the $xz$ band is
\begin{eqnarray}\label{eq:SigDeriv}
\Sigma_{xz}({\bm k},\omega) &=& - \frac{\sqrt{2}}{\pi r_{\rm s}} \int \! d^2{\bm q} \int_{-\infty}^{\infty} \! \frac{d \Omega}{2 \pi i} W_{xz}({\bm q},\Omega)\nonumber\\
&\times & G^{(0)}_{xz}({\bm k} + {\bm q},\omega + \Omega)~,
\end{eqnarray}
where we are now using dimensionless frequencies $(\omega,\Omega)$ and wavevectors $({\bm q},{\bm k})$ by expressing them in units of $ 2\varepsilon_{{\rm F}xz}/\hbar $ and $k_{{\rm F}xz}$, respectively. We define $r_{\rm s}$ here as in Eq.~(\ref{eq:defineRS}). The dimensionless Green's function is
\begin{equation}
G^{(0)}_{xz}({\bm k} + {\bm q},\omega + \Omega) = \frac{1}{\omega + \Omega - \left(\frac{\vert{\bm k} + {\bm q}\vert^2}{2} + \bar{\Delta}\right) + i \eta} 
\end{equation}
where the limit $\eta \rightarrow 0^{+}$ is implied for $\vert{\bm k} + {\bm q}\vert > k_{{\rm F}xz}$ and  $\eta \rightarrow 0^{-}$ for $\vert{\bm k} + {\bm q}\vert < k_{{\rm F}xz}$. Here $\bar{\Delta}$ is the band offset between the $xz$ and $yz$ bands and the $xy$ band in units of $2\varepsilon_{{\rm F}xz}$. Recent tight-binding calculations~\cite{guru_prb_2012} of SrTiO$_3$ heterostructures suggest that the spatial separation $d$ becomes small at large electron densities. In the limit $k_{{\rm F}xz}d \ll 1$, the dimensionless RPA screened interaction of Eq.~\ref{eq:RPAinteraction} reduces to
\begin{equation}\label{eq:ScreenedCoulomb}
\frac{1}{q + \sqrt{2} r_{\rm s}\left[\zeta^{-1/2}\chi^{(0)}_{xy}(\lambda q,\lambda^2\Omega) + \chi^{(0)}_{yz}(q,\Omega) +  \chi^{(0)}_{xz}(q,\Omega) \right]}
\end{equation}
where we have defined $\lambda = k_{{\rm F}xz}/k_{{\rm F}xy}$  and $\zeta = m_{\rm H}/m_{\rm L} $ has been introduced earlier. The density-response functions, $\chi^{(0)}_{\alpha}$, are dimensionless here and are obtained from the dimensionful functions~\cite{Giuliani_and_Vignale} by dividing by the negative density-of-states of the $\alpha$'th band.

\begin{figure}[t]
\includegraphics[width=1.0\linewidth]{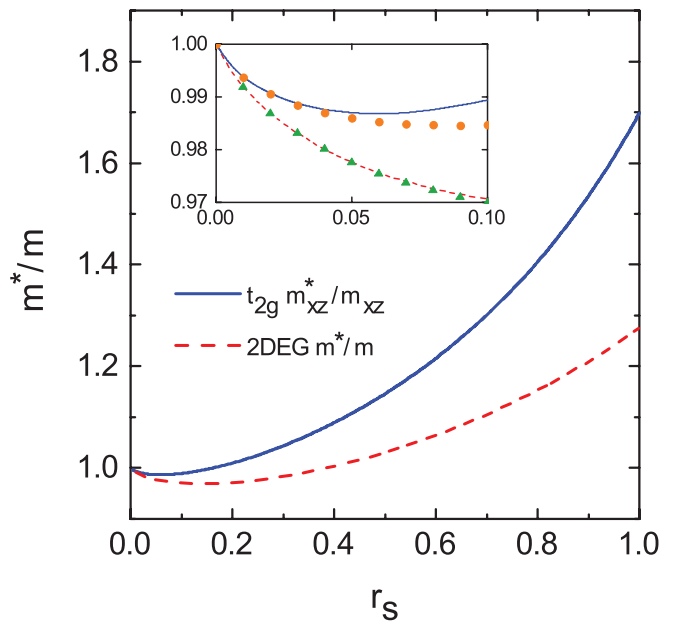}
\caption{(Color online)  The leading-order-in-$r_{\rm s}$ quasiparticle mass enhancement factor ({\it i.e.~}the inverse of Eq.~(\ref{eq:lowrs_effmass})) is plotted for $r_{\rm s} < 1$. The dashed red line shows the result for a isotropic single-band 2DEG, and the solid blue line shows the result for the $xz$-band of the $t_{2g}$ 2DEG, which also includes the effect of screening by electrons in the $yz$ and $xy$ bands. The Thomas-Fermi screened electron-electron interaction gives the leading-order contribution to quasiparticle mass enhancement at small $r_{\rm s}$ (high density). The presence of multiple filled bands in the $t_{2g} $ 2DEG yields a larger Thomas-Fermi screening wavevector, and therefore a larger quasiparticle mass enhancement than in a single-band 2DEG.\label{fig:eight}}
\end{figure}

Since the kinetic energy scales as $1/r^2_{\rm s}$ and the Coulomb interaction energy scales as $1/r_{\rm s}$, electron gases are weakly interacting in the high density (small $r_{\rm s}$) limit. We are then justified in ignoring the self-consistent nature of Eq.~(\ref{eq:QP_EE}) for the quasiparticle excitation energy and we can apply the ``on-shell'' approximation
\begin{equation}
\mathcal{E}_{xz}(k) \simeq \varepsilon_{xz}(k) + {\rm Re}\Sigma_{xz}(k,\varepsilon_{xz}(k)/\hbar) - \mu~.
\end{equation}
The inverse of the $xz$ band's quasiparticle mass enhancement factor is then given by
\begin{equation}\label{eq:eff_mass_dimless}
\frac{m_{xz}}{m^{\ast}_{xz}} = 1 + \frac{r^2_{\rm s}}{4}\left[ \frac{\partial}{\partial k} {\rm Re}\Sigma_{xz}(k,\varepsilon_{xz}(k)) \right]_{k=1}~.
\end{equation} 
To evaluate the wavevector derivative of the self-energy we need the derivative of the Green's function. Evaluating this on the Fermi surface we find
\begin{equation}
\begin{array}{l}\label{eq:kderiv_GF}
{\displaystyle \left.\frac{\partial}{\partial k}G^{(0)}_{xz}({\bm k} + {\bm q},\varepsilon(k) + \Omega) \right|_{k=1} }\vspace{0.1 cm} \\
{\displaystyle \qquad \qquad = \; -2 \pi i ~\delta(1-\vert \hat{\bm k} + {\bm q}\vert) \, \delta(\Omega) \,\left[(\hat{\bm k} + {\bm q}) \cdot \hat{\bm k} \right]}\vspace{0.1 cm} \\
{\displaystyle \qquad \qquad + \; (\hat{\bm k} \cdot {\bm q}) \left[ G^{(0)}_{xz}(\hat{\bm k} + {\bm q}, \varepsilon(1) + \Omega) \right]^2}~,
\end{array}
\end{equation}
where $\hat{\bm k} = {\bm k}/k$ is a unit vector. The first term on the right-hand side of Eq.~(\ref{eq:kderiv_GF}) gives the leading order in $r_{\rm s}$ contribution to the quasiparticle mass. The remaining integrations in Eq.~(\ref{eq:SigDeriv}) can be performed analytically.  After substituting the result into Eq.~(\ref{eq:eff_mass_dimless}), we obtain the inverse quasiparticle 
mass to $\mathcal{O}(r_{\rm s})$:
\begin{equation}\label{eq:lowrs_effmass}
\frac{m_{xz}}{m^{\ast}_{xz}} = 1 - \frac{r_{\rm s}}{\pi \sqrt{2}} \ln{\left( \frac{r_{\rm s} \delta}{2 \sqrt{2}} \right)} - \frac{\sqrt{2} \, r_{\rm s}}{\pi}
\end{equation}
where $\delta \equiv (\zeta^{-1/2} + 2) $. The expression appropriate to a single-band 2DEG is found by taking the limit $\delta \rightarrow 1$, which we have confirmed by comparing with full $G_0W$-RPA numerical calculations. The effective mass enhancement factor, which is given by the inverse of Eq.~(\ref{eq:lowrs_effmass}), is shown in Fig.~\ref{fig:eight}. 

As the inset to Fig.~\ref{fig:eight} shows, the small $r_{\rm s}$ expression (\ref{eq:lowrs_effmass}) is only quantitatively useful for $r_{\rm s} \lesssim 0.1$. It does, however, offer useful qualitative insight into the role of screening on the quasiparticle mass.  Physically, the renormalization of the electron mass derives from screened exchange scattering processes (see Fig.~\ref{fig:Feynman}) between the quasiparticle whose mass is under investigation, and other quasiparticles in the Fermi sea. Because of the Dirac delta functions in Eq.~(\ref{eq:kderiv_GF}), the leading order contribution comes from scattering amongst quasiparticles restricted to the Fermi surface, just as in the three-dimensional electron gas~\cite{DuBois,Rice}. For this reason, the presence of electrons in the $xy$ and $yz$ bands enters Eq.~(\ref{eq:lowrs_effmass}) only through their band-resolved density-of-states evaluated at the Fermi surface. The $xy$ band's finite density-of-states contributes the factor $\zeta^{-1/2}$ in the definition of $\delta$, while the $yz$ band's density-of-states contributes half of the factor of 2 in the definition of $\delta$ (the other half comes from screening by the $xz$ band's own electrons).

If we identify the above derivation as a simple application of first-order perturbation theory, but with the unscreened Coulomb interaction replaced with a Thomas-Fermi (TF) screened interaction~\cite{Giuliani_and_Vignale} (to which it is indeed equivalent), then we can understand in simple terms how increased screening acts to increase the quasiparticle mass. First consider that we know the exchange contribution to the self-energy tends to reduce the effective mass (in fact, the exchange level self-energy yields quasiparticles with \emph{zero} effective mass), so if a increase in the screening wavevector acts to reduce the exchange energy, then it will also have the effect of enhancing the quasiparticle mass. The unscreened exchange self-energy in Eq.~(\ref{eq:Sigma_ex}) for an electron of wavevector ${\bm k}$ is a sum over occupied states ${\bm q}$ of the bare amplitude $v_{\vert {\bm k}-{\bm q}  \vert } = 2 \pi e^2/(\kappa \vert {\bm k}-{\bm q} \vert)$. When we include a TF screening wavevector $\kappa_{\rm TF}$ (which is proportional to the total density-of-states at the Fermi surface), the bare amplitude is replaced by $v_{\vert {\bm k}-{\bm q}  \vert } \to 2 \pi e^2/(\kappa \vert {\bm k}-{\bm q} \vert + \kappa \kappa_{\rm TF}) $, which is smaller for all values of ${\bm q}$ to be summed over. Indeed, the larger $\kappa_{\rm TF} $ is (more specifically, the more electrons present at the Fermi surface), the smaller the TF self-energy. It is in this way that the $xy$ and $yz$ band's electrons act to increase the $xz$ band's quasiparticle masses.
\subsection{The role of anisotropy}
\label{Sect:Anisotropy}
\begin{figure}[t]
\centering
\includegraphics[width=1.0\linewidth]{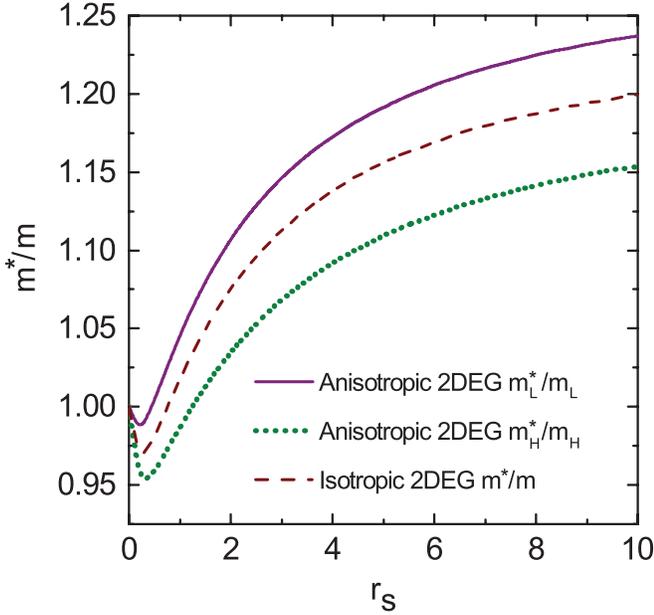}
\caption{(Color online) The quasiparticle mass enhancement factor of a isotropic single-band 2DEG (brown dashed line) compared against the light (purple) and heavy (green) quasiparticle mass enhancemnet factors of an anisotropic single-band 2DEG. \label{fig:sevenB}}
\end{figure}

Before presenting our numerical results for the $t_{2g}$ 2DEG, let us illustrate the general effect of band anisotropy on the quasiparticle mass in a simpler case. To seperate out this effect, we consider a single-band anisotropic 2DEG with the following non-interacting band structure
\begin{equation}\label{eq:NIdispersion}
\varepsilon({\bm k}) = \frac{\hbar^2 k^2_x}{2 m_{\rm L}} + \frac{\hbar^2 k^2_y}{2 m_{\rm H}}.
\end{equation}
We define the heavy and light quasiparticle mass factors $m^{\ast}_{\rm H}/m_{\rm H} $ and $m^{\ast}_{\rm L}/m_{\rm L} $ as above, and plot them in Fig.~\ref{fig:sevenB}  against $r_{\rm s}$. Here $r_{\rm s}$ is defined as in Eq.~(\ref{eq:defineRS}) but with $n_{xz} $ replaced by the total density of the single band. We also plot the quasiparticle mass factor $m^{\ast}/m$ for an isotropic two-dimensional electron gas for comparison. We find that $m^{\ast}_{\rm L}/m_{\rm L}$ ($m^{\ast}_{\rm H}/m_{\rm H}$) is enhanced (reduced) compared to the isotropic 2DEG mass enhancement factor $m^{\ast}/m$ at all values of $r_{\rm s}$, or equivalently, total 2DEG density.

To understand these results it is again helpful to consider the small $r_{\rm s}$ regime where the effective mass enhancement is governed by the TF screened self-energy. 
Consider first a single isotropic band with dispersion $\varepsilon(k)=\hbar^2{\bm k}^2/(2m_{\rm DOS})$ where as defined above $m_{\rm DOS}=\sqrt{m_{\rm H}m_{\rm L}}$. The isotropic 2DEG mass enhancement factor in the TF approximation can be written as
\begin{equation} \label{eq:TF_mass}
\frac{m^{\ast}_{\rm DOS}}{m_{\rm DOS}} = \frac{1}{1+\delta v^{\ast}_{\rm F}/v_{\rm F}}~,
\end{equation}
where $v_{\rm F}=\hbar k_{\rm F}/m_{\rm DOS}$ is the non-interacting band velocity at the Fermi energy, 
and the interaction contribution to the renormalized quasiparticle velocity at the Fermi energy is given by
\begin{equation}
 \delta v^{\ast}_{\rm F} = v^{\ast}_{\rm F} - v_{\rm F} = \hbar^{-1}\left[\partial_{k} \Sigma^{\rm TF}({\bm k})\right]_{{\bm k}=k_{\rm F}}~.
\end{equation}
The TF self-energy of a single-band isotropic 2DEG is given by 
\begin{eqnarray} \label{eq:Sigma_TF}
\Sigma^{\rm TF}({\bm k})&=& -\frac{1}{(2 \pi)^2}\int d^2{\bm q} \frac{2 \pi e^2}{\kappa \left(\vert{\bm k}-{\bm q} \vert + \kappa_{\rm TF}\right)}\nonumber\\
&\times&\Theta\left\{\varepsilon_{{\rm F}} - \varepsilon(\bm q)\right\}
\end{eqnarray} 
where the TF screening wavevector is defined
\begin{equation}\label{eq:TF_wavevector}
\kappa_{\rm TF}=\frac{2 \pi e^2}{\kappa}N(0)~.
\end{equation}
Here $N(0)$ is the total density-of-states at the Fermi surface. The quasiparticle mass enhancement factor calculated in this way is shown in Fig.~\ref{fig:eight}. Next consider keeping the density ($r_{\rm s}$) constant while introducing anisotropy in this single-band 2DEG by slowly deforming the shape of the Fermi surface from a circle to an ellipse with semimajor (semiminor) axes $k_{{\rm F} y}$ ($k_{{\rm F} x}$). The non-interacting dispersion of the anisotropic band is given by $\varepsilon({\bm k})=\hbar^2k^2_x/(2m_{\rm L}) + \hbar^2k^2_y /(2m_{\rm H})$ and the TF approximation for the light quasiparticle mass enhancement factor $m^{\ast}_{\rm L}/m_{\rm L} $ can be found from Eq.~(\ref{eq:TF_mass}) by replacing $v_{\rm F} = \hbar k_{\rm F}/m_{\rm DOS}$ with $v_{{\rm F} x} = \hbar k_{{\rm F} x}/m_{\rm L}$, and also replacing $ \delta v^{\ast}_{\rm F} =\hbar^{-1}\left[\partial_{k} 
\Sigma^{\rm TF}({\bm k})\right]_{{\bm k}=k_{\rm F}} $ with  $ \delta v^{\ast}_{{\rm F} x} =\hbar^{-1}\left[\partial_{k_x} \Sigma^{\rm TF}({\bm k})\right]_{{\bm k}=k_{{\rm F}x}} $:
\begin{equation} \label{eq:TF_mass_light}
\frac{m^{\ast}_{\rm L}}{m_{\rm L}} = \frac{1}{1+\delta v^{\ast}_{{\rm F}x}/v_{{\rm F}x}}~.
\end{equation}

By examining separately how the introduction of band anisotropy changes $\delta v^{\ast}_{{\rm F}x}$ 
relative to $\delta v_{\rm F}^{\ast}$, and $v_{{\rm F}x} $ relative to $v_{\rm F}$, we can identify the most important factor leading to $m^{\ast}_{\rm L}/m_{\rm L} > m^{\ast}_{\rm DOS}/m_{\rm DOS}$ for a given 
$r_{\rm s}$. Because the quasiparticle state at the Fermi surface in the light mass direction ($k_x = k_{{\rm F} x},\, k_y=0$) has more occupied states near it in momentum space when anisotropy is present, the TF self-energy of this state is increased in magnitude. Furthermore, since the TF self-energy acts to increase the quasiparticle velocity ({\it i.e.} decrease the quasiparticle mass, as shown in the previous section), it follows that $\delta v^{\ast}_{{\rm F}x} > \delta v^{\ast}_{\rm F} $. Thus electron-electron interactions tend to reduce $m^{\ast}_{\rm L}/m_{\rm L}$ below $m_{\rm DOS}^{\ast}/m_{\rm DOS}$ for a given value of $r_{\rm s}$. Despite this, the opposite occurs in Fig.~\ref{fig:sevenB}. The reason is that the band velocity $v_{{\rm F} x} = \hbar k_{{\rm F}x}/m_{\rm L}$ also enters the expression for the quasiparticle mass enhancement factor $m^{\ast}_{\rm L}/m_{\rm L}$, and for a fixed density we find $v_{{\rm F} x}>v_{\rm F}$, which tends to enhance $m^{\ast}_{\rm L}/m_{\rm L}$ above $m_{\rm DOS}^{\ast}/m_{\rm DOS}$. This latter effect is larger, and so while the change in quasiparticle velocity in the light mass direction is increased by anisotropy, the accompanying increase in band velocity is larger and results in $m^{\ast}_{\rm L}/m_{\rm L} > m_{\rm DOS}^{\ast}/m_{\rm DOS}$ for a given $r_{\rm s}$. Analogous reasoning leads to the conclusion that $m^{\ast}_{\rm H}/m_{\rm H} < m_{\rm DOS}^{\ast}/m_{\rm DOS}$ for a given $r_{\rm s}$. Although analysis based on the TF self-energy can only be gauranteed to hold at small $r_{\rm s}$, the conclusions it gives clearly persist within the full $G_0W$-RPA results of Fig.~\ref{fig:sevenB} to large $r_{\rm s}$.

\subsection{The $t_{2g}$ Quasiparticle Mass}
\begin{figure}[t]
\includegraphics[width=1.0\linewidth]{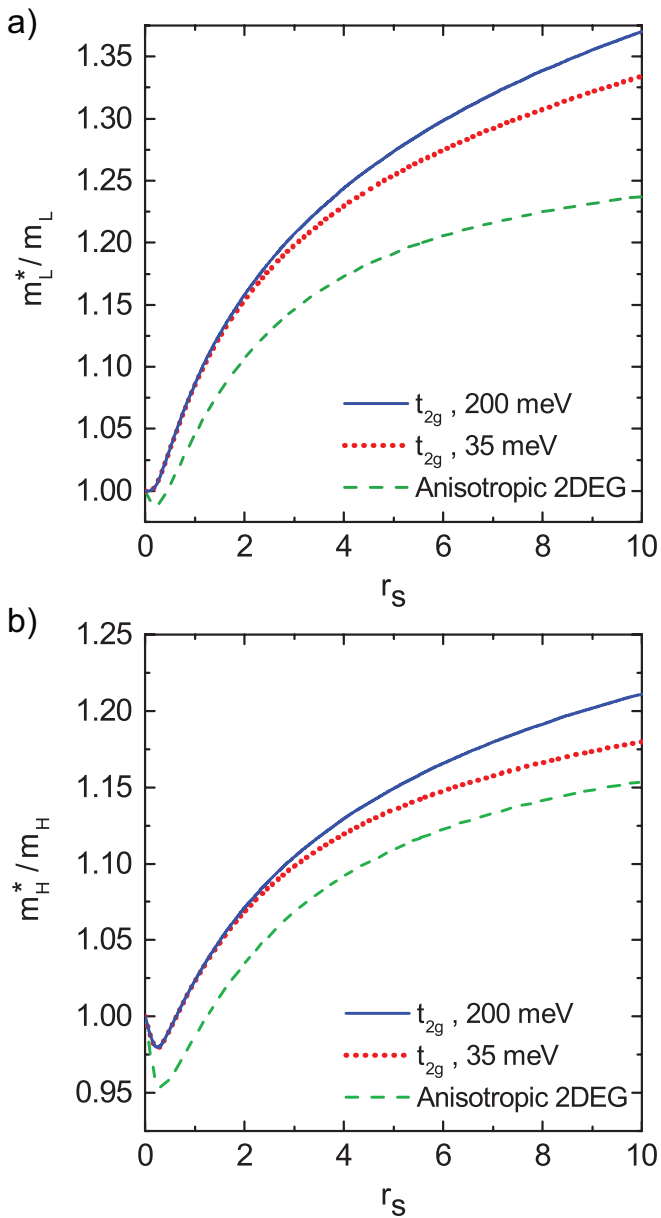}
\caption{(Color online) Panel (a) shows the light quasiparticle mass of the $t_{2g}$ 2DEG's $xz$ band plotted against $r_{s}$ for several values of the band offset $\Delta$ and separation $d$. For the red curve $\Delta = 35~{\rm meV}$ and $d=10a$ where $a=3.9$~\AA is the SrTiO$_3$ lattice constant. For the blue curve $\Delta = 200~{\rm meV}$ and $d=2a$.  The dashed green curve is for a single-band anisotropic 2DEG whose $\hat{\bm x}$-direction and $\hat{\bm y}$-direction non-interacting masses are the same as the $xz$ band in the $t_{2g}$ model. For this curve $r_{\rm s}$ is defined from the total density in the single anisotropic band. Panel (b) is the same, but for the heavy effective mass. \label{fig:masses}}
\end{figure}

In this section we present the results of our numerical calculation of the $G_0W$-RPA effective masses $m^{\ast}_{\rm H}$ and $m^{\ast}_{\rm L}$ of the $xz$ anisotropic band in the $t_{2g}$ 2DEG model. The results for the $yz$ band are identical because of symmetry. In Fig.~\ref{fig:masses} we compare the quasiparticle masses in the $t_{2g}$ 2DEG against those in a single-band anisotropic 2DEG with a non-interacting energy dispersion as in Eq.~(\ref{eq:NIdispersion}). Tight-binding studies of the $t_{2g}$ 2DEG created at SrTiO$_3$ surfaces or heterojunctions reveal that as the confining potential increases, so does the energy offset parameter $\Delta$, while the distance seperating the $xy$ band from the $xz$ and $yz$ bands $d$ decreases~\cite{guru_prb_2012}. We have included in Fig.~\ref{fig:masses} results for the $t_{2g}$ quasiparticle masses in the case of the very highest $t_{2g}$ electron densities ($\Delta = 200$ meV and $d=2a$ where $a=3.9$~\AA) as well as for the lowest $t_{2g}$ electron densities ($\Delta = 35$ meV and $d=10a$) at which we might practically neglect the presence of spin-orbit coupling in SrTiO$_3$ and thus reliably apply the $t_{2g}$ 2DEG model introduced in Sect.~\ref{Sect:Model}. In agreement with our analytic results in Section~{\ref{Sect:Analytic}, the quasiparticle masses in the $t_{2g}$ 2DEG are appreciably larger than their counterparts in the single-band anisotropic 2DEG. 

We were able to show in Sect.~\ref{Sect:Analytic} that at small $r_{\rm s}$ the increased quasiparticle mass in the anisotropic bands of the $t_{2g}$ 2DEG followed from a suppression of the wavevector derivative of the quasiparticle self-energy, which itself followed from an increase in the electronic screening of interactions due to the presence of electrons in the other bands (i.e.~the $xy$ and $yz$ bands). In that calculation we discovered that, at leading order in $r_{\rm s}$, the quasiparticle mass comes from exchange scattering via a reduced interaction in the form of the TF screened Coulomb interaction. Thus at small $r_{\rm s}$, because we knew that the additional electrons in the $xy$ and $yz$ bands would increase the TF screening wavevector, we knew the quasiparticle mass would be enhanced in the $t_{2g}$ 2DEG. In confirmation of this derivation, our numerical calculations reveal a substantial reduction in the wavevector derivative of the $xz$ bands self-energy when other bands are occupied. This reduction increases for increasing $r_{\rm s}$. At larger values of $r_{\rm s}$, however, correlation effects become important. Indeed, Eqs.~(\ref{eq:mLeq}) and~(\ref{eq:mHeq}), which define the quasiparticle masses, depend on the frequency derivative of the self-energy as well. Note that the self-energy is frequency independent in Hartree-Fock, and thus frequency dependence manifestly represents a correlation effect. Furthermore, we note that the frequency dependence of the self-energy does not enter the lowest order in $r_{\rm s}$ expressions for the quasiparticle mass derived in Sect.~\ref{Sect:Analytic}, as in this limit interactions are weak compared to the kinetic energy, and first-order perturbation theory is here equivalent to Hartree-Fock. Numerically, we find that the frequency derivative of the $xz$ bands quasiparticle self-energy is also suppressed by the presence of electrons in the $xy$ and $yz$ bands. Since the frequency and wavevector derivatives both are suppressed, some degree of cancellation occurs in Eqs.~(\ref{eq:mLeq}) and~(\ref{eq:mHeq}) for the $t_{2g}$ quasiparticle masses. Despite this, we find that the quasiparticle masses of the $t_{2g}$ anisotropic bands are enhanced from $25$ to $75$ percent above the quasiparticle mass values in a single-band anisotropic 2DEG. 

\section{Summary And Discussion}
\label{Sect:Summary}
Motivated by the recent synthesis of transition-metal oxide two-dimensional electron gases, we have introduced a model for studying the effects of many-body interactions in these systems. Using information from recent tight-binding~\cite{guru_prb_2012,millis_prb_2013} and \emph{ab initio} calculations~\cite{ghosez_prl_2011,guru_prb_2013,Popovic_prl_2008} we have chosen model parameters to specifically describe the electron gas formed in SrTiO$_3$. Our model captures the presence of band anisotropy, energy offsets between bands, and variable band confinement at the interface, all of which are characteristics likely to be shared by any two-dimensional electron gases formed from d-orbitals with anisotropic nearest-neighbor hopping amplitudes. Because the average conduction electron occupation number per transition-metal site is much less than one, the full long-range Coulomb interaction must be retained in any realistic interacting-electron model.  Our approach satisfies this criterion.  

We have used the $G_0W$-RPA approximation to calculate the self-energy contribution to the quasiparticle energy of the $xz$ and $yz$ anisotropic bands of the $t_{2g}$ two-dimensional electron gas. Because these bands' constituent d-orbitals have a large hopping amplitude in one direction in-plane and a small hopping amplitude in the other, the two-dimensional Fermi surfaces of these bands are approximately elliptical. Because rotational symmetry is broken, the self-energy depends on the  quasiparticle wavevector's orientation in momentum space, and the Fermi surface shape can be renormalized by interactions. By comparing the degree of Fermi surface renormalization in the anisotropic bands of the $t_{2g}$ model to the single-band anisotropic two-dimensional electron gas, we identified the reduction in band anisotropy as a rather universal effect, likely to occur in any anisotropic electron gas with long-range Coulomb interactions. 

Next we studied the impact of Coulomb interactions on the quasiparticle masses of the anisotropic bands in the $t_{2g}$ 2DEG model. We derived an analytic expression for the high-density (small $r_{\rm s}$) effective mass in both the $t_{2g}$ electron gas and the ordinary single-band isotropic two-dimensional electron gas. This leading order in $r_{\rm s}$ contribution to the two-dimensional quasiparticle mass was found to arise from exchange scattering via a reduced electron-electron interaction in the form of a TF screened Coulomb potential. The presence of multiple bands in the $t_{2g}$ case increased the TF screening wavevector, which substantially increased the quasiparticle masses $m^{\ast}_{\rm H}$ and $m^{\ast}_{\rm L}$. Numerical calculations at larger values of $r_{\rm s}$ confirm that the additional screening present in the $t_{2g}$ system from the multiple occupied bands increases the quasiparticle mass by reducing the wavevector dependence of the self-energy.

While the degree of Fermi surface shape renormalization is small and perhaps difficult to observe experimentally, the quasiparticle mass of the $t_{2g}$ anisotropic bands shows a large enhancement over the values expected in single-component 2DEGs. Shubnikov-de Haas oscillations are sensitive to the quasiparticle mass~\cite{Fang_and_Stiles,pudalov_prl_2002,Tan_prl_2005,Giuliani_and_Vignale}, but to our knowledge no clear signatures of the anisotropic band's Fermi surfaces have been reported in SrTiO$_3$ two-dimensional electron gases~\cite{kim_prl_2011}. When the mass is large, Landau-level spacing is small. It may be so small that disorder in current samples make oscillations attributable to the anisotropic bands undetectable. Perhaps the large quasiparticle mass we have found here helps to explain the lack of detection.
\begin{acknowledgments}
J.R.T. thanks the Scuola Normale Superiore di Pisa for their kind hospitality during part of this work. Work in Austin was supported by the DOE Division of Materials Sciences and Engineering under grant DE-FG02-ER45118. 
\end{acknowledgments}
\appendix

\section{Details of the Analytical calculation of FSSM in the $t_{2g}$ 2DEG}
In this appendix we outline the derivation of Eq.~(\ref{eq:ddkSigma_2DEG_four}) and Eq.~(\ref{eq:ddkSigma_2DEG_five}) from the main text. Let us begin with the former. We start using $k_{{\rm F}}$ to define dimensionless wavevectors and $\hbar k^2_{{\rm F}} /m_{\rm DOS}$ to define dimensionless frequencies. After expanding the wavevector derivative of the Green's function appearing in the $G_0W$-RPA expression for $\left.\partial_k\Sigma(k,0)\right|_{k=k_{\rm F}}$ to leading-order in the small-parameter $\cos{(\theta)}/q$ we obtain 
\begin{equation}\label{eq:APP_1}
\begin{array}{l}
{\displaystyle \left.\partial_k\Sigma(k,0)\right|_{k=k_{\rm F}} = \frac{-2 a_{\rm B} {\rm Ry}}{\pi^2 \kappa}\int^{\infty}_{0}\! q dq \int^{\pi}_{-\pi}\!d\theta \int^{\infty}_{0}\!d\Omega }\vspace{0.1 cm}\\
{\displaystyle  \quad \times \left(\frac{q^4-4\Omega^2}{\left\{q^4 + 4 \Omega^2\right\}^2}\right) \left(\frac{1/q}{1+\frac{\sqrt{2} r_{\rm s}}{q}\left\{ \frac{m_{\rm L}}{m_{\rm DOS}} \chi^{(0)}_{xy}(q,i\Omega)\right\}}\right)}
\end{array}
\end{equation}
where we have only written terms which will contribute at leading-order in powers of $1/r_{\rm s}$ in the final expression. The factor of $m_{\rm L}/m_{\rm DOS}$ appears in front of $\chi^{(0)}_{xy}(q,i\Omega)$ because the $xy$ band has a smaller band-mass and therefore a smaller density-of-states compared to the $xz$ and $yz$ bands. For simplicity we have set the confinement separation distance to zero, $d=0$, in the RPA screened interaction. We now rewrite the wavevector derivative of the self-energy using $k_{{\rm F}xy}$ and $\hbar k^2_{{\rm F}xy}/m_{\rm DOS} $ to define dimensionless wavevectors and frequencies, respectively. After this transformation it becomes clear that Eq.~(\ref{eq:APP_1}) appears to scale like $1/r_{\rm s}$. After introducing $R_{\rm s}$ and changing variables to $v=\Omega/q$ we find
\begin{widetext}
\begin{equation}\label{eq:APP_2}
 \left.\partial_k\Sigma(k,0)\right|_{k=k_{\rm F}} =\frac{-4 m_{\rm DOS} \, a_{\rm B} \,{\rm Ry}}{\pi \kappa m_{\rm L}} \, \frac{R_{\rm s}}{r_{\rm s}}\int^{\infty}_{0} dq  \int^{\infty}_{0} dv  \left(\frac{1/q}{1+\frac{\sqrt{2}R_{\rm s}}{q} \chi^{(0)}_{xy}(q,i q v)}\right)\left(\frac{q^2-4 \left(\frac{m_{\rm DOS}}{m_{\rm L}}\right)^2 v^2}{\left\{q^2 + 4\left(\frac{m_{\rm DOS}}{m_{\rm L}}\right)^2 v^2\right\}^2}\right)
\end{equation}
\end{widetext}
In the limit of $R_{\rm s} \ll 1$ we can approximate the Lindhard function along the imaginary frequency axis by its long-wavelength limit
\begin{equation}\label{eq:APP_3}
\lim_{q \rightarrow 0}\chi^{(0)}_{xy}(q,i q v) =1-\frac{v}{\sqrt{v^2+1}}~.
\end{equation}
When Eq.~(\ref{eq:APP_3}) is inserted into Eq.~(\ref{eq:APP_2}), we find that the integral diverges like $1/v$ in the long-wavelength limit. This is not suprising considering we have taken the large $\Omega$ and $q$ limit in obtaining Eq.~(\ref{eq:APP_2}). The analytic properties of the Lindhard function provide us with a convenient small $v$ ({\it i.e.} low energy) cutoff, $v_{c}$. Specifically, the Lindhard function is non-analytic~\cite{Giuliani_and_Vignale} in the sense that the long-wavelength limit is different depending on whether $v>1$ or $v<1$. Careful examination of the Lindhard functions for each band of the $t_{2g}$ 2DEG indicates that for $v<1$, screening from the elliptical $xz$ and $yz$ bands is important for convergence. Inclusion of these functions, however, leads to higher-order expressions in powers of $1/r_{\rm s}$. For $v > 1$ meanwhile, screening from the elliptic bands can be completely neglected. Applying the low-energy cutoff $v_{c}=1$, the remaining integrals can be completed to yield Eq.~(\ref{eq:ddkSigma_2DEG_four}) in the main text. 

The derivation of Eq.~(\ref{eq:ddkSigma_2DEG_five}) from the main text proceeds along very similar steps, which we briefly outline now.  Again we start by using $k_{{\rm F}}$ to define dimensionless wavevectors and $\hbar k^2_{{\rm F}} /m_{\rm DOS}$ to define dimensionless frequencies. After applying the coordinate transformation $k_x \rightarrow \zeta^{-1/4}k_x  $, $q_x \rightarrow \zeta^{-1/4}q_x$ , $k_y \rightarrow \zeta^{1/4}k_y$ and  $k_y \rightarrow \zeta^{1/4}k_y$, we find
\begin{widetext}
\begin{equation}
\begin{array}{l}\label{eq:APP_4}
{\displaystyle \Sigma_{xz}(k_{{\rm F}y},0) - \Sigma_{xz}(k_{{\rm F}x},0) = \frac{-2^{3/2} \bar{m}_{\rm DOS} {\rm Ry}}{\pi^2 \kappa^2 r_{\rm s}} \int^{\infty}_{0}\!dq \int^{\pi}_{-\pi}\!d\theta \int^{\infty}_{0}\!d\Omega  \left( \frac{1}{\gamma_{yz}+\frac{\sqrt{2} r_{\rm s} m_{\rm L}}{q m_{\rm DOS}} \chi_{xy}(\gamma_{yz}q,i\Omega)} \right) }\vspace{0.1 cm}\\
{\displaystyle \qquad \qquad \qquad \qquad \qquad \qquad \times \left(\frac{2 q \sin{(\theta)} - q^2}{4 \Omega^2 + \left\{2 q \sin{(\theta)} - q^2 \right\}^2} - \frac{2 q \cos{(\theta)} - q^2}{4 \Omega^2 + \left\{2 q \cos{(\theta)} - q^2 \right\}^2}  \right)~,}
\end{array}
\end{equation}
where we define $\gamma_{xz}=\sqrt{\zeta^{1/2}\cos^2{(\theta)} + \zeta^{-1/2}\sin^2{(\theta)}} $ and $\gamma_{yz}=\sqrt{\zeta^{-1/2}\cos^2{(\theta)} + \zeta^{1/2}\sin^2{(\theta)}} $. After some simple algebraic manipulations, and expanding to order $x^2$ in the small parameter $x=\cos{(\theta)}/q$ (which is the first non-vanishing term in the expansion) we obtain
\begin{equation}
\begin{array}{l}\label{eq:APP_5}
{\displaystyle  \Sigma_{xz}(k_{{\rm F}y},0) - \Sigma_{xz}(k_{{\rm F}x},0)  = \frac{8\sqrt{2}\bar{m}^3_{\rm DOS}{\rm Ry}}{\pi^2 \kappa^2 \bar{m}^2_{\rm L} } \frac{R_{\rm s}}{r^2_{\rm s}} \int^{\infty}_{0} \!dq \int^{\pi}_{-\pi}\!d\theta \int^{\infty}_{0}\!d\Omega  \, \frac{q^8-12q^4\left(\frac{m_{\rm DOS}}{m_{\rm L}}\right)^2\Omega^2  }{\left(q^4+4 \left(\frac{m_{\rm DOS}}{m_{\rm L}}\right)^2 \Omega^2 \right)^3} \, \cos^2{(\theta)} }\vspace{0.1 cm}\\
{\displaystyle \qquad \qquad \qquad \qquad \qquad \qquad \times  \left(\frac{1}{\gamma_{xz}+\frac{\sqrt{2}R_{\rm s}}{q} \chi_{xy}(\gamma_{xz}q,i\Omega)} - \frac{1}{\gamma_{yz}+\frac{\sqrt{2}R_{\rm s}}{q} \chi_{xy}(\gamma_{yz}q,i\Omega)}\right) ~,}
\end{array}
\end{equation}
where we are now using $k_{{\rm F}xy}$ and  $\hbar k^2_{{\rm F}xy}/m_{\rm DOS} $ to define dimensionless wavevectors and frequencies, respectively. For $R_{\rm s} \ll 1$ the leading-order term comes from $v>1$ again. With this low-energy cutoff, Eq.~(\ref{eq:APP_5}) remains convergent even when $R_{\rm s}=0$ in the denominator of the integrand. This allows us to evaluate the remaining integrals analytically and we finally obtain Eq.~(\ref{eq:ddkSigma_2DEG_five}) from the main text. The function ${\cal F}(\zeta)$ in Eq.~(\ref{eq:ddkSigma_2DEG_five}) is given by 
\begin{equation}\label{eq:APP_6}
 {\cal F}(\zeta) =   \frac{(1+\zeta) \, \left(\sqrt{\zeta}K\left[1-\zeta\right] + K\left[\frac{\zeta-1}{\zeta}\right]\right)}{(\zeta-1) \, \zeta^{1/4}} - \frac{(1+\zeta) \, \left(2\sqrt{\zeta}E\left[1-\zeta\right] + 2 \zeta E \left[\frac{\zeta-1}{\zeta} \right] \right)}{(\zeta-1) \,\zeta^{1/4}}
\end{equation}
where $K\left[x\right]$ and $E\left[x\right]$ are complete elliptic integrals of the first and second kind, respectively.
\end{widetext}

\begin{thebibliography}{99}
%
%
\bibitem{Hubbard}
	J Hubbard, 
	\href{http://dx.doi.org/10.1098/rspa.1963.0204}
	{Proc. R. Soc. Lond. A~{\bf 276}, 238 (1963)}.
%
\bibitem{Mattheiss}
	L.F. Mattheiss, 
	\href{http://dx.doi.org/10.1103/PhysRevB.6.4718}
	{Phys. Rev. B.~{\bf 6}, 4718 (1972)}.
%
\bibitem{stemmer_apl_2013}
        A.P. Kajdos, D.G. Ouellette, T.A. Cain and S. Stemmer
	\href{http://dx.doi.org/10.1063/1.4819203}
	{Appl. Phys. Lett.~{\bf 103}, 082120 (2013)}.
%
\bibitem{hwang_apl_2010}
        Y. Kozuka, M. Kim, H. Ohta, Y. Hikita, C. Bell, and H.Y. Hwang, 
        \href{http://dx.doi.org/10.1063/1.3524198}
	{Appl. Phys. Lett.~{\bf 97}, 222115 (2010)}.
%
\bibitem{mannhart_science_1010}
	J. Mannhart and D.G. Schlom, 
	\href{http://dx.doi.org/10.1126/science.1181862}{Science~{\bf 327}, 1607 (2010)}.
%
\bibitem{stemmer_armr_2014}
	S. Stemmer and S.J. Allen, 
	\href{http://dx.doi.org/10.1146/annurev-matsci-070813-113552}
	{Annu. Rev. Mater. Res.~{\bf 44}, 151 (2014)}.
%
\bibitem{levy_armr_2014}
	J.A. Sulpizio, S. Ilani, P. Irvin, and J. Levy, 
	\href{http://dx.doi.org/10.1146/annurev-matsci-070813-113437}
	{Annu. Rev. Mater. Res.~{\bf 44}, 117 (2014)}.
%
\bibitem{stemmer_prb_2012}
	P. Moetakef, C.A. Jackson, J. Hwang, L. Balents, S.J. Allen, and S. Stemmer, 
	\href{http://dx.doi.org/10.1103/PhysRevB.86.201102}
	{Phys. Rev. B.~{\bf 86}, 201102(R) (2012)}.
%
\bibitem{balents_prb_2013}
	R. Chen, S. Lee, and L. Balents, 
	\href{http://dx.doi.org/10.1103/PhysRevB.87.161119}
	{Phys. Rev. B.~{\bf 87}, 161119(R) (2013)}.
%
\bibitem{Bohm_and_Pines}
	D. Bohm and D. Pines, 
	\href{http://dx.doi.org/10.1103/PhysRev.92.609}
	{Phys. Rev.~{\bf 92}, 609 (1953)}.
%
\bibitem{jonson}
	M. Jonson, \href{http://dx.doi.org/10.1088/0022-3719/9/16/012}
	{J. Phys. C: Solid State Phys.~{\bf 9}, 3055 (1976)}.
%
\bibitem{stern_prl_1967}
        F. Stern,
	\href{http://dx.doi.org/10.1103/PhysRevLett.18.546}
	{Phys. Rev. Lett.~{\bf 18}, 546 (1967)}.
%
\bibitem{grecu}
	D. Grecu, \href{http://dx.doi.org/10.1088/0022-3719/8/16/014}
	{J. Phys. C: Solid State Phys.~{\bf 8}, 2627 (1975)}.
%
\bibitem{vinter_prl_1975}
        B. Vinter,
	\href{http://dx.doi.org/10.1103/PhysRevLett.35.1044}
	{Phys. Rev. Lett. {\bf 35}, 1044 (1975)}.
%
\bibitem{santoro_prb_1989}
        G.E. Santoro and G.F. Giuliani,
	\href{http://dx.doi.org/10.1103/PhysRevB.39.12818}
	{Phys. Rev. B.~{\bf 39}, 12818 (1989)}.
%
\bibitem{Asgari_PRB_2005}
	R. Asgari, B. Davoudi, M. Polini, G.F. Giuliani, M.P. Tosi, and G. Vignale, 
	\href{http://dx.doi.org/10.1103/PhysRevB.71.045323}
	{Phys. Rev. B.~{\bf 71}, 045323 (2005)}.
%
\bibitem{polini_prb_2008}
        M. Polini, R. Asgari, G. Borghi, Y. Barlas, T. Pereg-Barnea and A. H. MacDonald,
	\href{http://dx.doi.org/10.1103/PhysRevB.77.081411}
	{Phys. Rev. B.~{\bf 77}, 081411(R) (2008)}.
%
\bibitem{bostwick_science_2010}
	A. Bostwick, F. Speck, T. Seyller, K. Horn, M. Polini, R. Asgari, A. H. MacDonald and E. Rotenberg, 
	\href{http://dx.doi.org/10.1126/science.1186489}
	{Science~{\bf 328}, 999 (2010)}.
%
\bibitem{Pines_and_Nozieres}
	D. Pines and P. Nozi\'eres, {\it The Theory of Quantum Liquids} (W.A. Benjamin, Inc., New York, 1966).
%
\bibitem{Giuliani_and_Vignale}
	G.F. Giuliani and G. Vignale, {\it Quantum Theory of the Electron Liquid} 
	(Cambridge University Press, Cambridge, 2005).
%
\bibitem{guru_prb_2012}
	G. Khalsa and A.H. MacDonald, 
	\href{http://dx.doi.org/10.1103/PhysRevB.86.125121}
	{Phys. Rev. B.~{\bf 86}, 125121 (2012)}.
%
\bibitem{millis_prb_2013}
	S.E. Park and A.J. Millis, 
	\href{http://dx.doi.org/10.1103/PhysRevB.87.205145}
	{Phys. Rev. B.~{\bf 87}, 205145 (2013)}.
%
\bibitem{guru_prb_2013}
	G. Khalsa, B. Lee, and A.H. MacDonald, 
	\href{http://dx.doi.org/10.1103/PhysRevB.88.041302}
	{Phys. Rev. B.~{\bf 88}, 041302 (2013)}.
%
\bibitem{Popovic_prl_2008}
	Z.S. Popovic, S. Satpathy, and R.M. Martin,
	\href{http://dx.doi.org/10.1103/PhysRevLett.101.256801}
	{Phys. Rev. Lett.~{\bf 101}, 256801 (2005)}.
%
\bibitem{ghosez_prl_2011}
        P. Delugas, A. Filippetti, V. Fiorentini, D.I. Bilc, D. Fontaine, and P. Ghosez,
	\href{http://dx.doi.org/10.1103/PhysRevLett.106.166807}
	{Phys. Rev. Lett.~{\bf 106}, 166807 (2011)}.
%
\bibitem{Quinn_and_Ferrell}
        J.J. Quinn and R.A. Ferrell, 
	\href{http://dx.doi.org/10.1103/PhysRev.112.812}
	{Phys. Rev.~{\bf 112}, 812 (1958)}.
%
\bibitem{Goodenough}
	J.B. Goodenough, {\it Localized to Itinerant Electronic Transitions in Perovskite Oxides} 
	(Springer, Berlin, 1996).
%
\bibitem{Allen_prb_2013}
        S.J. Allen, B. Jalan, S. Lee, D.G. Ouellette, G. Khalsa, J. Jaroszynski, 
        S. Stemmer, and A.H. MacDonald,
	\href{http://dx.doi.org/10.1103/PhysRevB.88.045114}
	{Phys. Rev. B.~{\bf 88}, 045114 (2013)}.
%
\bibitem{Rotenberg_prb_2010}
	Y.J. Chang, A. Bostwick, Y.S. Kim, K. Horn, and E. Rotenberg, 
	\href{http://dx.doi.org/10.1103/PhysRevB.81.235109}
	{Phys. Rev. B.~{\bf 81}, 235109 (2010)}.
%
\bibitem{king_natcomm_2014}
	P.D.C. King, S. M. Walker, A. Tamai, A. de la Torre, T. Eknapakul, P. Buaphet, 
	S.-K. Mo, W. Meevasana, M. S. Bahramy, and F. Baumberger, 
	\href{http://dx.doi.org/10.1038/ncomms4414}
	{Nature Commun.~{\bf 5}, 3414 (2014)}.
%
\bibitem{ilani_natcomm_2014}
	A. Joshua, S. Pecker, J. Ruhman, E. Altman, and S. Ilani, 
	\href{http://dx.doi.org/10.1038/ncomms2116}
	{Nature Commun.~{\bf 3}, 1129 (2012)}.
%
\bibitem{Cancellieri_prb_2014}
	C. Cancellieri, M.L. Reinle-Schmitt, M. Kobayashi, V.N. Strocov, 
	P.R. Willmott, D. Fontaine, P. Ghosez, P. Delugas, and V. Fiorentini, 
	\href{http://dx.doi.org/10.1103/PhysRevB.89.121412}
	{Phys. Rev. B.~{\bf 89}, 121412 (2014)}.
%
\bibitem{Comment}
	Although the long-wavelength low-temperature static dielectric constant of bulk SrTiO$_3$ 
	is in the tens of thousands [A.S. Barker Jr. and M. Tinkham, 
	\href{http://dx.doi.org/10.1103/PhysRev.125.1527}
	{Phys. Rev. {\bf 125}, 1527 (1962)}] because of the 
	presence of a soft LO phonon mode near the $\Gamma$ point, 
	the effective dielectric constant which screens electron-electron interactions is expected to be substantially 
	smaller. The electronic transitions which contribute to the self-energy in the $t_{2g}$ model are on the scale
	of a few hundred meV, much larger than the soft-phonon mode energy which is closer 
	to a few meV [Y. Yamada and G. Shirane, 
	\href{http://dx.doi.org/10.1143/JPSJ.26.396}{J. Phys. Soc. Jpn. {\bf 26}, 396 (1969)}]. 
	Screening by this mode is therefore weak. 
	Furthermore, the LO phonon mode is soft only in close vicinity to the $\Gamma$ point. 
	Even static electric fields are ineffectively screened by this mode unless they are constant 
	over distances which greatly exceed a lattice constant. Based on these considerations, 
	we think that the effective dielectric constant to be included in electron-electron interaction 
	calculations is closer to $\kappa \sim 15$, similar to small-gap covalent semiconductors. 
	All of our results are presented in terms of the parameter $r_{\rm s}$ 
	and thus are independent of the value of $\kappa$. 
	The conversion from $r_{\rm s}$ to density, however, does depend on $\kappa$. At liquid helium temperatures the dielectric constant of bulk SrTiO$_3$ is $\kappa \sim 10$ in the range $2-40$ meV and $\kappa \sim 7$ above $40$ meV [R.C. Neville, B. Hoeneisen and C.A. Mead, 
	\href{http://dx.doi.org/10.1063/1.1661463}{J. Appl, Phys. {\bf 43}, 2124 (1972)}].  

%
\bibitem{Fang_and_Stiles}
	F.F. Fang and P.J. Stiles, 
	\href{http://dx.doi.org/10.1103/PhysRev.174.823}
	{Phys. Rev.~{\bf 174}, 823 (1968)}.
%
\bibitem{pudalov_prl_2002}
	V.M. Pudalov, M.E. Gershenson, H. Kojima, N. Butch, E.M. Dizhur, G. Brunthaler, A. Prinz, 
	and G. Bauer, \href{http://dx.doi.org/10.1103/PhysRevLett.88.196404}
	{Phys. Rev. Lett.~{\bf 88}, 196404 (2002)}.
%
\bibitem{Tan_prl_2005}
	Y.-W. Tan, J. Zhu, H.L. Stormer, L.N. Pfeiffer, K.W. Baldwin, and K.W. West,
	\href{http://dx.doi.org/10.1103/PhysRevLett.94.016405}
	{Phys. Rev. Lett.~{\bf 94}, 016405 (2005)}.
%
\bibitem{dassarma_prb_2004}
        S. Das Sarma, V.M. Galitski, and Y. Zhang,
	\href{http://dx.doi.org/10.1103/PhysRevB.69.125334}
	{Phys. Rev. B.~{\bf 69}, 125334 (2004)}.
%
\bibitem{mahan_prl_1989}
        G.D. Mahan and B.E. Sernelius,
	\href{http://dx.doi.org/10.1103/PhysRevB.62.2718}
	{Phys. Rev. B.~{\bf 62}, 2718 (1989)}.
%
\bibitem{macdonald_prb_1994}
	L. Zheng and A.H. MacDonald, 
	\href{http://dx.doi.org/10.1103/PhysRevB.49.5522}
	{Phys. Rev. B.~{\bf 49}, 5522 (1994)}.
%
\bibitem{hedin_pr_1965}
	L. Hedin, 
	\href{http://dx.doi.org/10.1103/PhysRev.139.A79}
	{Phys. Rev.~{\bf 139}, A79 (1965)}.
%
\bibitem{Fetter_and_Walecka}
	A.L. Fetter and J.D. Walecka, {\it Quantum Theory of Many-Particle Systems} 
	(McGraw-Hill, 1971).
%
\bibitem{Luttinger}
        J. M. Luttinger, 
	\href{http://dx.doi.org/10.1103/PhysRev.119.1153}
	{Phys. Rev. {\bf 119}, 1153 (1960)}.
%
\bibitem{DuBois}
        D.F. DuBois, \href{http://dx.doi.org/10.1016/0003-4916(59)90062-4}{Ann. Phys.~{\bf 8}, 24 (1959)}.
%
\bibitem{Rice}
        T.M. Rice, \href{http://dx.doi.org/10.1016/0003-4916(65)90234-4}{Ann. Phys.~{\bf 31}, 100 (1965)}.
%
\bibitem{kim_prl_2011}
	M. Kim, C. Bell, Y. Kozuka, M. Hikita, and H.Y. Hwang, 
	\href{http://dx.doi.org/10.1103/PhysRevLett.107.106801}
	{Phys. Rev. Lett. {\bf 107}, 106801 (2011)}.
%
\end{thebibliography}
\end{document}